\pdfoutput=1
\documentclass[preprint,nofootinbib,amsmath,amssymb,aps,prd,floatfix]{revtex4-2}
\usepackage{graphicx,mathrsfs,color}
\usepackage[caption=false]{subfig}
\usepackage[colorlinks=true,allcolors=blue]{hyperref}

\newcommand{\beq}{\begin{eqnarray}}
\newcommand{\eeq}{\end{eqnarray}}
\newcommand{\non}{\nonumber\\}

\newcommand{\p}{\partial}

\DeclareMathOperator{\tr}{tr}

\DeclareMathOperator{\SU}{SU}
\DeclareMathOperator{\SO}{SO}

\DeclareMathOperator{\MeV}{MeV}
\DeclareMathOperator{\fm}{fm}

\newcommand{\bn}{\mathbf{n}}

\newcommand{\bx}{\mathbf{x}}

\newcommand{\btau}{\boldsymbol{\tau}}
\newcommand{\bpi}{\boldsymbol{\pi}}

\newcommand{\Lag}{\mathcal{L}}

\renewcommand{\i}{\mathrm{i}}
\renewcommand{\d}{\mathop{}\!\mathrm{d}}

\newcommand{\calE}{\mathcal{E}}

\begin{document}

\title{Nonlinear rigid-body quantization of Skyrmions}
\author{Sven Bjarke Gudnason}
\email{gudnason(at)henu.edu.cn}
\affiliation{Institute of Contemporary Mathematics, School of
  Mathematics and Statistics, Henan University, Kaifeng, Henan 475004,
  P.~R.~China}
\begin{abstract}
  We consider rigid-body quantization of the Skyrmion in the most
  general four-derivative generalization of the Skyrme model with a
  potential giving pions a mass, as well as in a class of higher-order
  Skyrme models.
  We quantize the spin and isospin zeromodes following the results of
  Pottinger and Rathske.
  Although one could hope that a one-parameter family of theories
  could provide a smaller spin contribution to the energy at some point
  in theory space -- which would be welcome for BPS-type models, we
  find that the standard Skyrme model
  limit, with two time
  derivatives, gives rise to the smallest spin contribution to the
  energy.
  We speculate whether this tuning of the spin energy could be useful
  in the larger picture of quantizing vibrational and light massive
  modes of the Skyrmions.
  Finally, we establish a topological energy bound for the
  Pottinger-Rathske model with potential terms as well as new bounds
  for higher-order Skyrme models, with and without a potential.
\end{abstract}
\maketitle
\newpage

\section{Introduction}

The Skyrme model \cite{Skyrme:1961vq,Skyrme:1962vh} is a low-energy effective field theory
description of 
QCD in a pure pion theory, where baryons are solitons known as
Skyrmions.
Although the model was proposed already in the sixties by Skyrme, it
first received serious attention after Witten showed that the Skyrmion
is the nucleon of QCD in the large-$N_c$ limit in the seminal papers
\cite{Witten:1983tw,Witten:1983tx}.
Although the Skyrme model provides a qualitative description of the
nucleon to about the 30\% level of accuracy compared with experiments
\cite{Adkins:1983ya}, a major obstacle in using the theory for nuclei is that it
gets the binding energies wrong by roughly an order of magnitude,
already at the classical level.
The community has worked on solving this problem at the classical
level, essentially by finding so-called BPS limits of the theory, for
which there exist solutions\footnote{There exists a Bogomol'nyi bound
for the standard Skyrme model, but there are no solutions saturating
the bound, which means all solutions have positive binding energy. },
the idea being that once one have found the BPS limit, a small
perturbation could create the tiny binding energies of about 1\% of
the nucleon mass per baryon.
To list a few of the attempts to find a BPS-type model for Skyrmions, the
Sutcliffe model \cite{Sutcliffe:2010et,Sutcliffe:2011ig} is a
five-dimensional Yang-Mills theory that realizes 
a flat-space holographic description of Skyrmions as the holonomy of
the gauge fields \cite{Atiyah:1989dq}; the BPS-Skyrme model is a radical change of
the Skyrme model by eliminating the model and replacing it by the
topological charge current squared as well as a suitable potential
\cite{Adam:2010fg,Adam:2010ds}; finally, the weakly-bound Skyrme model is based on an energy
bound using the H\"older inequality for which the Skyrme term and a
potential to the fourth power saturates the energy bound
\cite{Harland:2013rxa,Gillard:2015eia}. 
The Sutcliffe model requires an infinite number of vector bosons added
to the Skyrme model in order to reach the BPS limit, i.e.~the limit in
which the classical energy is directly proportional to the baryon
number, hence yielding vanishing classical binding energy. The
advantage of the model, similarly to holographic constructions \cite{Sakai:2004cn}
and the hidden local symmetry approach \cite{Harada:2003jx}, is that all the
couplings to the infinite tower of vector bosons are determined.
The BPS-Skyrme model has (analytic) solutions for any baryon number
that saturates the Bogomol'nyi bound, but near-BPS solutions turn out
to be a numerically challenging problem \cite{Gillard:2015eia,Gudnason:2020tps,Gudnason:2021gwc,Gudnason:2022aig}.
The weakly-coupled Skyrme model only saturates the Bogomol'nyi bound
for a single baryon, hence all nuclei already have a nonvanishing
albeit small classical binding energy. It turns out that all the
solutions take the shape of lattices of point-particle-like Skyrmions
\cite{Gillard:2015eia}.
For completeness, we can mention that a dielectric formulation of the
Skyrme model \cite{Adam:2020iye} also provides small binding energies \cite{Adam:2020iye,Gudnason:2020ftf}, but
also in this case the solutions tend to be point-particle-like
constellations like in the weakly-bound Skyrme model \cite{Gudnason:2020ftf}.

Now as put forward in the recent paper \cite{Gudnason:2023jpq}, although the BPS race
that has taken place for over 10 years in the community has given rise
to interesting ideas and some analytic solutions, the vanishing
classical binding energy does not solve the problem of the binding
energy of nuclei, simply due to the spin contribution.
In order to illustrate this, let's consider the standard Skyrme model
with the rational map approximation \cite{Houghton:1997kg,Battye:2001qn}:
\beq
E = \int\d^3x\;\left[\lambda(a_1+a_2B)\calE_2 + \lambda^{-1}(b_1+b_2B)B\calE_4\right],
\eeq
with $a_{1,2}$, $b_{1,2}$ being positive coefficients, $B$ the baryon
number and $\lambda$ the length scale.
Using the rational map approximation is a good approximation to
Skyrmions with baryon numbers $B=1,2,\ldots,7$ in the massive Skyrme
model, i.e.~once the pion mass term is turned on and overestimates the
energies only by a few percent
\cite{Battye:2004rw,Battye:2006tb,Gudnason:2022jkn}.\footnote{
A multi-layered rational map may be utilized for larger Skyrmions, where
two or more different angular maps are utilized in radial layers of
the soliton \cite{Manton:2000kj,Feist:2012ps}.
}
Finding the Derrick stability, we obtain
\beq
\lambda = \sqrt{\frac{e_4(b_1+b_2B)B}{e_2(a_1+a_2B)}},
\eeq
which for large $B$ goes like $\lambda\propto\sqrt{B}$.
The size of the Skyrmion hence grows like $\sqrt{B}$.
Since the mass of the Skyrmion grows at least as fast as $B$, the
moment of inertia scales like $B^2$ or higher.
The spin contribution to the energy found in the seminal paper by
Adkins-Nappi-Witten thus goes like
\beq
E_{\rm spin} = \frac{J^2}{2\Lambda} \propto \frac{J^2}{B^2}.
\eeq
Hence, even for nuclei whose ground state has a spin, the spin
contribution is suppressed by roughly $B^2$ and quickly becomes
negligible
(for $B=7$, the suppression is by a factor of $1/49$).
Even more troublesome are the nuclei that are bosonic with spin 0 and
isospin 0 in the ground state, as their contribution is just zero.
These nuclei exist for $B$ up to $40$ for stable nuclei and $48$ for
``long-lived'' nuclei.
For nuclei with $B\gtrsim40$, the Coulomb repulsion begins to be
important, so the isospin quantum number is generically nonvanishing
in the ground state of such nuclei.

Now let us contemplate a BPS model, which by definition has vanishing
classical binding energy for the Skyrmions.
Using only rigid-body quantization, we can thus compute the binding
energy of e.g.~${}^4$He as
\beq
\Delta = 4(M_1 + E_{\rm spin}) - (4M_1 + 0)
= \frac{3}{2\Lambda}.
\eeq
Computing this number within the standard Skyrme model as a rough
estimate, one obtains of the order of $4\times 4.6\%$ of the nucleon
mass, which is about $175\MeV$.
The physical binding energy of ${}^4$He is about $28.3\MeV$.
For nuclei with nonvanishing spin and/or isospin in the ground state,
the problem is, of course, slightly less severe.

We can thus see that the BPS models cannot solve the binding
energy problem of the Skyrme model in the scheme of rigid-body
quantization, as also explained in Ref.~\cite{Gudnason:2023jpq}.
In the latter reference, it was proposed that since the number of
zeromodes are fixed and independent of the baryon number\footnote{To
be more precise, the number of rotational and isorotation zeromodes
are 6 for $B>1$, whereas spin and isospin are equal in magnitude for
the $B=1$ Skyrmion due to spherical symmetry.}, the quantum
contribution is underestimated for $B>1$ nuclei and the resolution is
to take more modes into account in the quantization procedure.
We can now see that there are two ways to approach the problem:
in the spirit of Ref.~\cite{Gudnason:2023jpq} one could take as many degrees of
freedom as needed into account and quantize them to hopefully arrive at a
cancellation of contributions that land just right on the nuclear
physics scale of about $8\MeV$ per nucleon.
Alternatively, one could believe in the semi-classical approximation
of solitons being a good description of nature, with the classical
contribution (mass) being the dominant one, and all quantum
corrections being much smaller in magnitude, thus possibly avoiding too
large cancellations (fine tuning) in the final result.

With the latter notion of naturalness in mind, which is also
confirming the validity of using solitons in the first place, one may
ask whether it is possible to lower the spin contribution to the
nucleon.
In this paper, we study this question in a generalization of the
Skyrme model to the most general Lorentz-invariant Lagrangian written
in terms of the chiral Lagrangian field $U\in\SU(2)$ with up to four
derivatives and up to second order in a polynomial potential.
Unfortunately, our result is, as it turns out, the Skyrme model limit
of the model at hand gives the smallest spin contribution to the
energy for the nucleon.
For the naturalness path forward, one would thus stick with the Skyrme
term, whereas if one proceeded along the cancellation path to nuclear
phenomenology, this model introduces an extra parameter that could be
used to fine tune the binding energies.

The generalization of the Skyrme model to the most general
fourth-order derivative term, has been studied previously in the
literature in the context of the Skyrme model and the chiral
Lagrangian.
In the scheme of an EFT, there are only two different fourth-order
derivative terms involving the pion matrix $U$
\cite{Gasser:1983yg,Gasser:1984gg,Fearing:1994ga,Bijnens:1999sh,Bijnens:2022zqo}.\footnote{
One might naively think that one could also write down the term
$\tr(U^\dag\square U)^2$, but by using integration by parts and field
redefinitions, it can be shown to be equivalent to
the two fourth-order derivative terms studied in this paper as well as
a combination of some higher-than-fourth-order terms.
}
Scrutinizing $\pi\pi$ scattering data in the D-wave using the chiral
Lagrangian, Weinberg's renowned result \cite{Weinberg:1966kf} did not
apply, and Gasser and Leutwyler's result was that a nonvanishing
coefficient of the squared kinetic term was favored
\cite{Gasser:1983ky,Gasser:1983kx}.
With this result Donoghue, Golowich and Holstein used the Skyrme model
with this new kinetic term squared to predict the proton mass from the
$\pi\pi$ scattering data, obtaining $880\pm300\MeV$
\cite{Donoghue:1984yq}.
They calculated the contribution to the mass from the proton's spin by
using Adkins, Nappi and Witten's (ANW) leading order formula
\cite{Adkins:1983ya}, but with the full moment of inertia of the
Skyrmion \cite{Donoghue:1984yq} (see also
Ref.~\cite{Andrianov:1986dn}).
This was considered a good approximation, since the deviation from the
Skyrme model limit (i.e.~vanishing kinetic term squared) was
experimentally quite small.
The nucleon-nucleon potential, which turned out not to lead to an
attractive force at medium distances in the standard Skyrme model, was
computed for many generalizations in the search for this attractive
property and also in the Skyrme model with the kinetic term
squared \cite{Lacombe:1985yd,Lacombe:1985mr}.
They also did not find the exact contribution of the spin energy, but
simply computed the moment of inertia and used it in the ANW formula.

The exact computation of the spin contribution to the energy of the
Skyrmion was first done by Pottinger and Rathske
\cite{Pottinger:1985fc}, by solving the cubic equation relating the
spin operator squared to the moments of inertia using Cardano's
formula, which we will review.
Although Donoghue, Golowich and Holstein worked with the same model
earlier, we will here denote the massless Skyrme model with the
squared kinetic term as the Pottinger-Rathske (PR) model, since they
treated the spin contribution to the energy exactly and not
perturbatively, as other groups did.

The kinetic term squared, in contrast to the Skyrme term, contains
four time derivatives, which in turn makes it lose $\SO(4)$ Euclidean
symmetry in the Hamiltonian after performing the usual Legendre
transformation from the Lorentz-invariant Lagrangian.
It also gives the unpleasing side effect of opening up for run-away
directions in the Hamiltonian energy.
This is generically not unexpected for low-energy effective field
theories at higher orders in the derivative expansion.

The kinetic term squared has also been considered in further
literature, in the context of Skyrme-type models, see the review
\cite{Schwesinger:1988af}.
In particular, the nucleon-nucleon potential has been studied in 
an extension of the PR model with a sextic derivative
term, being the baryon current squared \cite{Oka:1987tg}.
Finite density computations of the energy in a hybrid model with both
quarks and pions have been considered, using also the kinetic term
squared \cite{Hahn:1987xr}.
Solitons with nonvanishing Hopf number were also studied in the Skyrme
model with the squared kinetic term \cite{Fujii:1984zp}.
The Skyrmion and in particular the Skyrme model with the squared
kinetic term can be related to the soliton in the Nambu-Jona-Lasinio
model by a derivative expansion \cite{Zuckert:1994st}.
The Skyrme model, including the squared kinetic term, was generalized
to include several sextic derivative terms and these terms were used
to improve the fitted value of the pion decay constant
\cite{Kumar:1999ij}.
The stability versus metastability aspects of the Skyrmion was ported
from QCD to electroweak Skyrmions in Ref.~\cite{Ellis:2012cs}.
A non-singular spacetime defect soliton has been studied on a
non-simply connected topology with nontrivial field solutions using
the Skyrme model with the squared kinetic term
\cite{Klinkhamer:2013mha}.
Closed time-like curves were studied in the Skyrme model with the
squared kinetic term -- all coupled to Einstein gravity
\cite{Sorba:2014rjf}.
In all the papers of this paragraph, either the perturbative ANW
formula for the spin contribution to the energy was used, or only the 
stability/metastability aspects of the Skyrmion were studied.

In this paper, we point out that the spin contribution to the energy
changes with a positive definite correction upon including an
arbitrary higher-order derivative term in the Lagrangian
that contains 4 time derivatives, under these
conditions: The higher-order derivative term is Lorentz invariant and
its static energy is positive definite.
This means that including 4 time derivatives in a Skyrme-type model,
instead of 2 time derivatives, can only increase the spin contribution
to the energy and hence exacerbate the binding energy problem.
We illustrate this claim by considering generalizations of the Skyrme
model with the kinetic term squared (the PR model) as
well as with 8th, 10th and 12th order derivative terms that contain
four time derivatives (and no d'Alembertian).

The paper is organized as follows.
In Sec.~\ref{sec:model} we review the Pottinger-Rathske (PR) model and
collective coordinate (or rigid-body) quantization therein.
In Sec.~\ref{sec:calib} we give our chosen calibration scheme and in
Sec.~\ref{sec:numerics} we present the numerical results of the
paper.
In Sec.~\ref{sec:higherorder} we include a class of higher-order
Skyrme models with four time derivatives and between 8 and 12
derivatives in total, whose quantization is identical to the model of
Sec.~\ref{sec:model} with modified moments of inertia.
We conclude in Sec.~\ref{sec:discussion} with a discussion and
outlook.
We have delegated technical details of the models in the paper to
appendices.
In particular, the positivity of the static energy of the
PR model is given in
App.~\ref{app:positivity_static_PR} and of the higher-order models in
App.~\ref{app:positivity_static_HO}.
A topological energy bound for the massless PR model is
reviewed in App.~\ref{app:topobound_PR} and bounds are found for
the case that includes non-derivative potentials.
Finally, new topological energy bounds are found for the higher-order
models in App.~\ref{app:topobound_HO}.

\section{The Pottinger-Rathske Skyrme model}\label{sec:model}

We consider the chiral Lagrangian with the most general
Lorentz-invariant Lagrangian, up to fourth order in derivatives
\cite{Gasser:1983yg,Gasser:1984gg,Fearing:1994ga,Bijnens:1999sh,Bijnens:2022zqo},
which was considered previously by Donoghue-Golowich-Holstein
\cite{Donoghue:1984yq} and by Pottinger-Rathske (PR)
\cite{Pottinger:1985fc}.
In addition to the derivative terms, we include the standard pion mass
term \cite{Adkins:1983hy}, as well as the loosely bound potential term
\cite{Gudnason:2016mms,Gudnason:2016cdo}; 
hence the total Lagrangian is also the most general Lagrangian up to
polynomials of order 2, in the chiral Lagrangian field, $U$
\cite{Gudnason:2016cdo}:
\begin{align}
  \Lag &= \frac{F_\pi^2}{16}\tr(R_\mu R^\mu)
  +\frac{1}{32e^2}\tr\left([R_\mu,R_\nu][R^\mu,R^\nu]\right)
  -\frac{\beta}{32}\left(\tr(R_\mu R^\mu)\right)^2\non
  &\phantom{=\ }
  -\frac{F_\pi^2m_\pi^2}{8}\tr(\mathbf{1}_2 - U )
  -\frac{F_\pi^2M^2}{32}\left[\tr(\mathbf{1}_2 - U)\right]^2,
  \label{eq:Lphysicalunits}
\end{align}
where the right-invariant chiral current is
\beq
R_\mu = \p_\mu U U^\dag,
\eeq
$F_\pi$ is the pion decay constant, $e$ is the Skyrme coupling
constant, $\beta$ is the dimensionless coupling of the other
fourth-order derivative term which we shall dub the kinetic term
squared, $m_\pi$ is the pion mass, $M$ is the mass parameter of the
loosely bound potential term, the chiral Lagrangian or Skyrme field
$U$ is related to the pions via 
\beq
U = \mathbf{1}_2\sigma + \i\bpi\cdot\btau,
\eeq
where $\btau$ are the standard Pauli spin matrices, and finally we
use the Minkowski metric with the mostly positive signature.

By the most general fourth-order derivative theory with only the field
$U$, we mean that field redefinitions and integration-by-parts
relations have been taken into account, leaving the two displayed
terms as a representation of the two independent terms that exist, at
this order in the derivative expansion
\cite{Gasser:1983yg,Gasser:1984gg,Fearing:1994ga,Bijnens:1999sh,Bijnens:2022zqo}. 

The topological charge of the field $U$ is known as the baryon number
and can be computed as
\beq
B = -\frac{1}{24\pi^2}\int\d^3x\;\epsilon^{ijk}\tr(R_i R_j R_k),
\label{eq:B}
\eeq
which arises due to the finite-energy condition on $U$ implying that
$\lim_{|x|\to\infty}U=\mathbf{1}_2$ (or another constant, that however
can be rotated into the unit matrix) which in turn effectively
point-compactifies 3-space to $S^3$ and hence
$\pi_3(\SU(2)_{\rm L}\times\SU(2)_{\rm R}/\SU(2)_{\rm diag})=\mathbb{Z}\ni B$.

We first redefine new dimensionless coupling constants as
\beq
\frac{1}{e^2} = \alpha^2(2\eta-1), \qquad
\beta = 2\alpha^2(1-\eta), \qquad
\eta\in[-1,1],
\eeq
where $\eta$ interpolates between the two fourth-order soliton-stabilizing
terms and $\alpha$ is an overall positive coefficient.
For the details of the positivity of the static energy in the PR
model, see App.~\ref{app:positivity_static_PR}.
The Lagrangian now reads
\begin{align}
  \Lag &= \frac12\tr(R_\mu R^\mu)
  +\frac{2\eta-1}{16}\tr\left([R_\mu,R_\nu][R^\mu,R^\nu]\right)
  -\frac{1-\eta}{8}\left(\tr(R_\mu R^\mu)\right)^2\non
  &\phantom{=\ }
  -m_1^2\tr(\mathbf{1}_2 - U )
  -\frac{m_2^2}{4}\left[\tr(\mathbf{1}_2 - U )\right]^2,
  \label{eq:L}
\end{align}
where the energy and length units are rescaled as
\beq
\mu = \frac{F_\pi\alpha}{4}, \qquad
\lambda = \frac{2\alpha}{F_\pi},
\label{eq:mu_lambda}
\eeq
and the dimensionless mass parameters are given by
\beq
m_1 := \frac{2\alpha m_\pi}{F_\pi}, \qquad
m_2 := \frac{2\alpha M}{F_\pi}.
\eeq
The real parameter $\eta$ interpolates between three different models,
see Tab.~\ref{tab:model_limits}.
\begin{table}[!ht]
\begin{center}
\begin{tabular}{l||c|c}
  & Skyrme term  & kinetic term squared\\
  \hline\hline
  $\eta=1$ & 1 & 0\\
  $\eta=1/2$ & 0 & $1/2$\\
  $\eta=-1$ & $-3$ & $2$
\end{tabular}
\caption{A one-parameter family interpolating between 3 different
  models, with $\eta=1$ being the Skyrme model limit. }
\label{tab:model_limits}
\end{center}
\end{table}

The Lagrangian splits into potential and kinetic energy as
\begin{align}
  L &= T^L - V, \\
  V &= \int\d^3x\bigg[
    -\frac12\tr(R_i^2)
    -\frac{2\eta-1}{16}\tr\left([R_i,R_j]^2\right)
    +\frac{1-\eta}{8}\left(\tr(R_i^2)\right)^2\non
      &\phantom{=\int\d^3x\bigg[\ }
    +m_1^2\tr(\mathbf{1}_2 - U)
    +\frac{m_2^2}{4}\left[\tr(\mathbf{1}_2 - U)\right]^2
  \bigg],\label{eq:V}\\
  T^L &= \int\d^3x\bigg[
    -\frac12\tr(R_0^2)
    -\frac{2\eta-1}{8}\tr\left([R_0,R_i]^2\right)
    -\frac{1-\eta}{8}\left(\tr(R_0^2)\right)^2
    +\frac{1-\eta}{4}\tr(R_0^2)\tr(R_i^2)
  \bigg].
\end{align}
Using the hedgehog Ansatz
\beq
U = \mathbf{1}_2\cos f(r) + \i\hat{\bx}\cdot\btau\sin f(r),
\label{eq:hedgehog}
\eeq
with $\hat{\bx}=\bx/r$ being a unit 3-vector in $\mathbb{R}^3$,
$r=|\bx|$ and the potential or static energy becomes
\begin{align}
  V &= \int\d^3x\bigg[
    (f')^2 + \frac{2\sin^2f}{r^2}
    +\frac{1-\eta}{2}(f')^4
    +2\eta\frac{\sin^2(f)(f')^2}{r^2}
    +\frac{\sin^4f}{r^4}\non
    &\phantom{=\int\d^3x\bigg[\ }
    +2m_1^2(1-\cos f)
    +m_2^2(1-\cos f)^2
    \bigg],
    \label{eq:Vf_simp}
\end{align}
which can be seen to be positive (semi-)definite, term by term, for
$\eta\in[0,1]$.
In order to see that this is still a positive static energy functional
for $\eta\in[-1,1]$, we rewrite it as
\begin{align}
  V &= \int\d^3x\bigg[
    (f')^2 + \frac{2\sin^2f}{r^2}
    +\frac{1+\eta}{2}\left(\frac{2\sin^2(f)(f')^2}{r^2} + \frac{\sin^4f}{r^4}\right)
    +\frac{1-\eta}{2}\left((f')^2 - \frac{\sin^2f}{r^2}\right)^2\non
    &\phantom{=\int\d^3x\bigg[\ }
    +2m_1^2(1-\cos f)
    +m_2^2(1-\cos f)^2
    \bigg],
    \label{eq:Vf_positive}
\end{align}
which is indeed positive semi-definite for $\eta\in[-1,1]$.

The static equation of motion for the profile function of the Skyrmion
is found by varying the static potential energy \eqref{eq:Vf_simp}
with respect to $f$, yielding:
\begin{align}
  f''
  +\frac{2f'}{r}
  -\frac{\sin2f}{r^2}
  +3(1-\eta)(f')^2f''
  +2(1-\eta)\frac{(f')^3}{r}
  +\frac{2\eta\sin^2(f)f''}{r^2}\phantom{=0\ }\non
  \mathop+\frac{\eta\sin(2f)(f')^2}{r^2}
  -\frac{\sin^2(f)\sin2f}{r^4}
  -m_1^2\sin f
  -m_2^2(1-\cos f)\sin f = 0,\label{eq:eomf}
\end{align}
which needs to be accompanied by the boundary conditions $f(0)=\pi$
and $f(\infty)=0$ corresponding to a unit-Skyrmion ($B=1$).

Introducing a classical rotation of the spherically symmetric hedgehog
Skyrmion can be done in two ways, either by isospinning the soliton
\beq
U^{(A)}(\bx,t) = A(t)U(\bx)A(t)^\dag,
\label{eq:AUAdag}
\eeq
which however is equivalent to spinning it via
\beq
U^{(A)}(\bx,t) = U(\mathsf{R}(t)\bx), \qquad
\mathsf{R}_{ij} = \frac12\tr(\tau^i A\tau^j A^\dag),
\eeq
where $U(\bx)$ is a static solution to the field equations.
The (classical) kinetic part of the Lagrangian can now readily be written down
\begin{align}
  T^L &= \int\d^3x\bigg[
    -\frac12\tr(T_i T_j)
    -\frac{2\eta-1}{8}\tr\left([T_i,R_k][T_j,R_k]\right)
    +\frac{1-\eta}{4}\tr(T_i T_j)\tr(R_k^2)
    \bigg]a_i a_j\non
  &\phantom{=\ }
  +\int\d^3x\bigg[
    -\frac{1-\eta}{8}\tr(T_i T_j)\tr(T_k T_l)
  \bigg]a_i a_j a_k a_l,
\end{align}
where $T_i=\frac{\i}{2}[\tau^i,U]U^\dag$ and the
$\mathfrak{su}(2)$-valued angular momenta, $a_i\tau^i$, are defined as
\beq
a_i = -\i\tr(\tau^iA^\dag\dot{A}).
\eeq
In general, the kinetic energy is quite a complicated expression;
however, for the spherically symmetric hedgehog Ansatz
\eqref{eq:hedgehog}, the integrals reduce to
\beq
T = J_ia_i - T^L
= \frac12\Lambda_1 a_i^2
- \frac34\Lambda_2 a_i^2 a_j^2,
\label{eq:T}
\eeq
with the momentum conjugate to $a_i$:
\beq
J_i = \frac{\p T^L}{\p a_i}
= \Lambda_1 a_i - \Lambda_2(a_j)^2 a_i.
\label{eq:J}
\eeq
and
\begin{align}
  \Lambda_1 &= \frac{16\pi}{3}\int\d r\; r^2\left[
  \sin^2f
  +\eta\sin^2(f)(f')^2
  +\frac{\sin^4f}{r^2}
  \right],\label{eq:Lambda1}\\
  \Lambda_2 &= (1-\eta)\frac{64\pi}{15}\int\d r\;r^2\sin^4f,
  \label{eq:Lambda2}
\end{align}
where we have used the following angular integrals
\beq
\int\d\theta\d\phi\;\sin\theta\hat{x}^i\hat{x}^j = \frac{4\pi}{3}\delta^{ij},\qquad
\int\d\theta\d\phi\;\sin\theta\hat{x}^i\hat{x}^j\hat{x}^k\hat{x}^l
= \frac{4\pi}{15}(\delta^{ij}\delta^{kl} + \delta^{ik}\delta^{jl} + \delta^{il}\delta^{jk}).\nonumber
\eeq
Every term in $\Lambda_1$ is positive definite for $\eta\in[0,1]$, but
this is not the case for $\eta$ in the full range of static
positivity, namely $\eta\in[-1,1]$.
It is expected that $\Lambda_1$ takes the minimum value for $\eta=-1$
and the maximum value for $\eta=1$ (ignoring the fact that the
solution $f$ depends on $\eta$), whereas the opposite is expected for
$\Lambda_2$. 
There are two competing effects at play: $\eta\to1$ (from below) increases the moment of
inertia by increasing $\Lambda_1$, but it also decreases $\Lambda_2$,
which reduces the moment of inertia by the quartic kinetic term, and
vice versa for $\eta\to-1$.

The basic strategy for quantizing the isospin rotation, which for
spherical symmetry is equal to a spatial rotation of the hedgehog
Skyrmion, is to write the kinetic energy \eqref{eq:T} in terms of
$J_iJ_i$, which is the square of the momentum conjugates.
If this is possible, canonical quantization applies and we simply
replace the operator $(2J)^2$ with $\frac{\p^2}{\p A_{ij}^2}$, with
$A_{ij}$ being the elements of the isospin rotation matrix $A$ of
Eq.~\eqref{eq:AUAdag}; the latter is just the Laplacian operator on
the 3-sphere \cite{Adkins:1983ya}, which has eigenvalues
$\ell(\ell+2)$, with $\ell=0,1,2,3,\ldots$, $\ell=2j$ and $j$ is the
isospin quantum number. 
As a result the lowest spin of a fermion $j=\tfrac12$ corresponds to
$\ell=1$ and hence $J^2=\tfrac34$.
Unfortunately, simply squaring Eq.~\eqref{eq:J} does not allow for the
elimination of $a_i$, so we need to solve for $a_i$ and insert the
result into Eq.~\eqref{eq:T}, which hopefully is a function of
$J_iJ_i$, i.e.~the momentum conjugate squared. 
Although it is difficult to compute $a_i$ in terms of $J_i$ of
Eq.~\eqref{eq:J}, the squared quantities are easier to handle:
\beq
J^2 = \Lambda_1^2 a^2 + \Lambda_2^2 (a^2)^3 - 2\Lambda_1\Lambda_2(a^2)^2.
\eeq
We can now invert the equation by treating the equation as a classical
cubic polynomial in $a^2$.
Before solving the equation, it will prove useful to analyze the
polynomial a bit before doing so.
Let us write
\beq
y(x) = \Lambda_2^2 x^3 - 2\Lambda_1\Lambda_2 x^2 + \Lambda_1^2 x - J^2,\qquad
x := a^2.
\eeq
We find the saddle points straightforwardly as $y'(x)=0$:
\beq
x_{\pm} = \left(\frac23 \pm \frac13\right)\frac{\Lambda_1}{\Lambda_2},
\eeq
which are both positive.
The values of the polynomial function $y(x_\pm)$ at the saddle points
determine the number and kind of roots the polynomial possesses.
In particular, we get
\beq
y(x_-) = \frac{4\Lambda_1^3}{27\Lambda_2} - J^2, \qquad
y(x_+) = -J^2.
\eeq
The first (in the $x$-direction) saddle point value can change sign
depending on the values of the integrals $\Lambda_1$ and $\Lambda_2$.
Since $\Lambda_1$ is an increasing function with $\eta$ and
$\Lambda_2$ is a decreasing function with $\eta$, the ratio
$\Lambda_1^3/\Lambda_2$ is an increasing function with $\eta$, that
diverges at $\eta=1$.
This means that for large $\eta\sim1$, $y(x_-)$ is positive and
$y(x_+)$ is negative.
In this case, there are 3 real roots of $y$, which is physically
puzzling.
It turns out that the 3 roots exists in most of the model's parameter
space, but only 1 of the roots gives a positive spin contribution to
the energy -- this root connects to the Skyrme model limit and we will
denote it the physical root \cite{Pottinger:1985fc}.

In order to reach the form for which Cardano's result applies, we
shift the variable as 
$x=\xi+\frac23\ell$, for which the
polynomial reads
\beq
\frac{y}{\Lambda_2^2} = \xi^3 - \frac13\ell^2\xi
+ \frac{2}{27}(\ell^3 - \varsigma^2),
\label{eq:root_equation}
\eeq
where we have defined
\beq
\ell = \frac{\Lambda_1}{\Lambda_2}, \qquad
\varsigma = \sqrt{\frac{27}{2}}\frac{J}{\Lambda_2},
\label{eq:ell_sigma}
\eeq
for which we have three roots of $y$ (assuming that $2\ell^3>\varsigma^2$) as
\beq
\xi_{0,\pm} =
\frac{2\ell}{3}\cos\left(\frac{\theta-\pi\pm2\pi}{3}\right), \qquad
\theta = \arccos\left(1 - \frac{\varsigma^2}{\ell^3}\right),
\label{eq:xi_theta}
\eeq
where the index $0$ means that $\pm2\pi$ in the cosine is replaced by $0$ and
\beq
x_{0,\pm} = \xi_{0,\pm} + \frac23\ell,
\eeq
with $x_-$ being the smallest positive root.

Writing the kinetic quantum energy in terms of $J^2$, we get
\beq
T &= \frac12\Lambda_1 x_a - \frac34\Lambda_2 x_a^2,
\label{eq:Tx}
\eeq
with $x_a$, $a=0,\pm$ being one of the roots.
Near the Skyrme model limit ($\eta$ close to unity), it is safe to
assume that $\varsigma^2\ll\ell^3$, since $\Lambda_1$ is maximal and
$\Lambda_2$ tends to zero.
In this case, we can expand in $1/\ell$, obtaining
\begin{align}
  T(x_-) &= \frac{J^2}{2\Lambda_1}
  + \frac{\Lambda_2J^4}{4\Lambda_1^4}
  + \frac{\Lambda_2^2J^6}{2\Lambda_1^7}
  + \frac{3\Lambda_2^3J^8}{2\Lambda_1^{10}}
  + \mathcal{O}\left(\frac{J^{10}\Lambda_2^4}{\Lambda_1^{13}}\right),\label{eq:Tx-}\\
  T(x_{0,+}) &= -\frac{\Lambda_1^2}{4\Lambda_2}
  \mp \sqrt{\frac{\Lambda_1}{\Lambda_2}}|J|
  - \frac{J^2}{4\Lambda_1}
  \pm \frac{\Lambda_2^{1/2}|J|^3}{8\Lambda_1^{5/2}}
  - \frac{\Lambda_2J^4}{8\Lambda_1^4}
  +\mathcal{O}\left(\frac{|J|^5\Lambda_2^{3/2}}{\Lambda_1^{11/2}}\right),
\end{align}
yielding the order-$\Lambda_2$ correction to the usual
Skyrme model's spin energy for the case of the smallest (physical) root,
$T(x_-)$, plus higher-order corrections.
Notice that $T(x_-)$ only increases once a nonvanishing
$\Lambda_2$ is introduced.
The only possibility of lowering the spin correction to the mass,
would be if the change in $\eta$ would increase $\Lambda_1$ much
faster than it would increase $\Lambda_2$.

The term $-\Lambda_1^2/4\Lambda_2$ in $T(x_{0,+})$ signals that the
two cases of $x_0$ and $x_+$ are unphysical; firstly these roots do
not connect to the Skyrme model limit and secondly they give a
negative contribution to the energy (that diverges in the Skyrme model
limit).
This fact and also the $N_c$ counting leads one to discard the two
roots $x_0$ and $x_+$, see Ref.~\cite{Pottinger:1985fc}.

Returning to the physical root, $x_-$, it will prove convenient for
numerical computations to rewrite the spin contribution to the energy
as
\beq
T = \frac{\Lambda_1^2}{3\Lambda_2}\cos\frac\theta3\left(1-\cos\frac\theta3\right),
\label{eq:Tphys}
\eeq
with $\theta$ given by Eq.~\eqref{eq:xi_theta}, which can also be
written as \cite{Pottinger:1985fc}
\beq
\theta = \arccos\left(1 - \frac{\varsigma^2}{\ell^3}\right)
=\arccos\left(1 - \frac{27\Lambda_2J^2}{2\Lambda_1^3}\right).
\eeq
The spin contribution \eqref{eq:Tphys} is positive definite and given in
Ref.~\cite{Pottinger:1985fc}, but it is more difficult to see whether
turning on $\Lambda_2$ increases (or decreases) the spin contribution,
which however can easily be seen from Eq.~\eqref{eq:Tx-}.

In order to restore the units, we have to substitute
$\Lambda_1\to\mu\lambda^2\Lambda_1$ and
$\Lambda_2\to\mu\lambda^4\Lambda_2$.
Inserting these into the spin contribution \eqref{eq:Tphys}, we obtain
\beq
T = \frac{\mu\Lambda_1^2}{3\Lambda_2}\cos\frac\theta3\left(1-\cos\frac\theta3\right),\qquad
\theta=\arccos\left(1 - \frac{27\hbar^2\Lambda_2J^2}{2\Lambda_1^3}\right),
\label{eq:Tphys_units}
\eeq
where we have set $\hbar:=1/(\mu\lambda)=2\alpha^{-2}$ and we have 
used the energy and length units of Eq.~\eqref{eq:mu_lambda}.
The spin contribution is thus directly proportional to the energy
unit, as one would expect and we can also see that the angle $\theta$
depends on $\hbar$ or rather the model parameter $\alpha$, which needs
to be calibrated.

\subsection{Calibration}\label{sec:calib}

In order to compare the quantum spin energy \eqref{eq:Tphys_units} to 
the classical mass of the Skyrmion $\mu V$, we must calibrate the
model.
The overall scale is simply $\mu$ (of Eq.~\eqref{eq:mu_lambda}) for
both energies, but their ratio depends also on $\alpha$, as can be
seen from Eq.~\eqref{eq:Tphys_units} where $\hbar=2\alpha^{-2}$. 
In order to fix $\alpha$, we need to perform a calibration that
entails fitting one energy quantity and one length scale, which we
will do next.

We choose to fit the nucleon mass ($M_N$) to the classical Skyrmion mass
$V$ and the size of the Skyrmion $R$ to the electric charge radius of
the nucleon ($R_N$), so we obtain the following equations
\beq
\mu V = M_N,\qquad
\lambda R = R_N,
\eeq
where $V$ is given by Eq.~\eqref{eq:V}, the radius is computed as a
weighted integral
\beq
R =
\sqrt{\frac{-\frac{1}{24\pi^2}\int\d^3x\;|x|^2\epsilon^{ijk}\tr(R_i R_j R_k)}{B}},
\eeq
and the energy scale $\mu$ and the length scale $\lambda$ are given in
Eq.~\eqref{eq:mu_lambda}.
We use the experimental values $939\MeV$ and $0.8783\fm$ for the
nucleon mass and radius, respectively.
Solving for the pion decay constant and $\alpha$, we have
\begin{align}
  F_\pi = \frac{2R}{R_N}\alpha,\qquad
  \alpha = \sqrt{\frac{2M_NR_N}{V R}}.
  \label{eq:Fpi_alpha}
\end{align}
We choose to fit to the classical Skyrmion mass, since including the spin
correction is complicated by the fact that the nontrivial and
nonlinear $\alpha$ dependence makes an analytic formula unavailable.
If one were to choose to fit the total Skyrmion mass to the physical
mass, only numerical methods would be able to do so.
Since we know that there are further corrections to the Skyrmion
energy, e.g.~from vibrational modes \cite{BjarkeGudnason:2018bju}, we
simplify the problem and fit just to the classical Skyrmion mass
here.

\subsection{Numerical results}\label{sec:numerics}

We solve the equation of motion \eqref{eq:eomf} for the static
Skyrmion, and use it to compute the two moments of inertia,
$\Lambda_{1,2}$ of Eqs.~\eqref{eq:Lambda1}-\eqref{eq:Lambda2}.
We use simple gradient flow as the numerical method.
Then we use Eq.~\eqref{eq:Fpi_alpha} to calibrate the model, which in
turn determines the energy and length scales \eqref{eq:mu_lambda}.
With these at hand, and using the appropriate spin of the nucleon
$j=1/2$ yielding $J^2=3/4$, we can compute the spin correction to the
energy using Eq.~\eqref{eq:Tphys_units} and other observables of the
model. 

\begin{figure}[!htp]
  \centering
  \mbox{\subfloat[]{\includegraphics[width=0.49\linewidth]{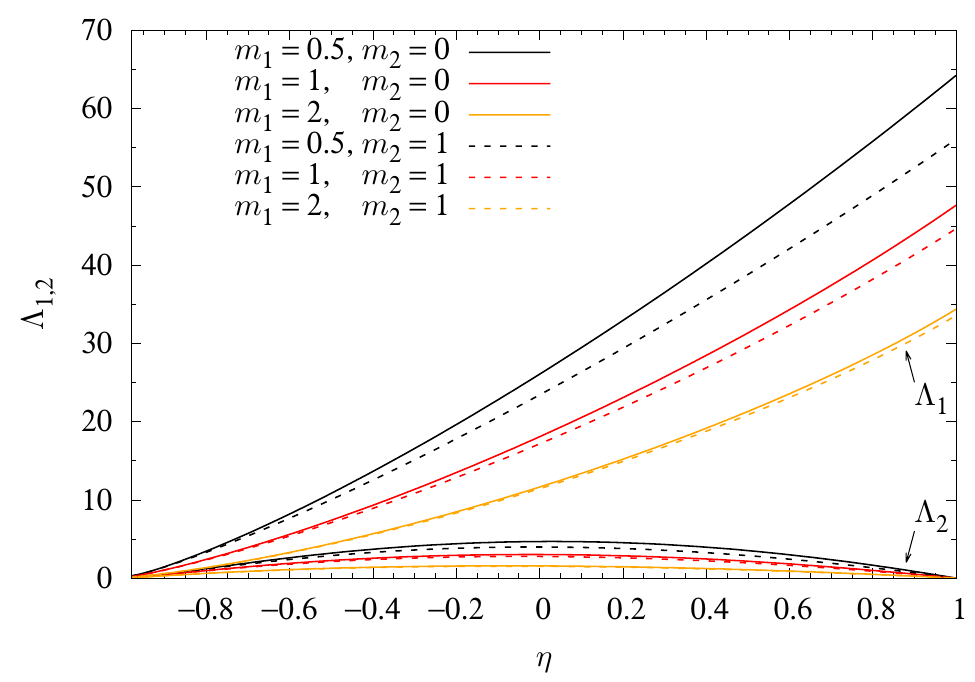}}
    \subfloat[]{\includegraphics[width=0.49\linewidth]{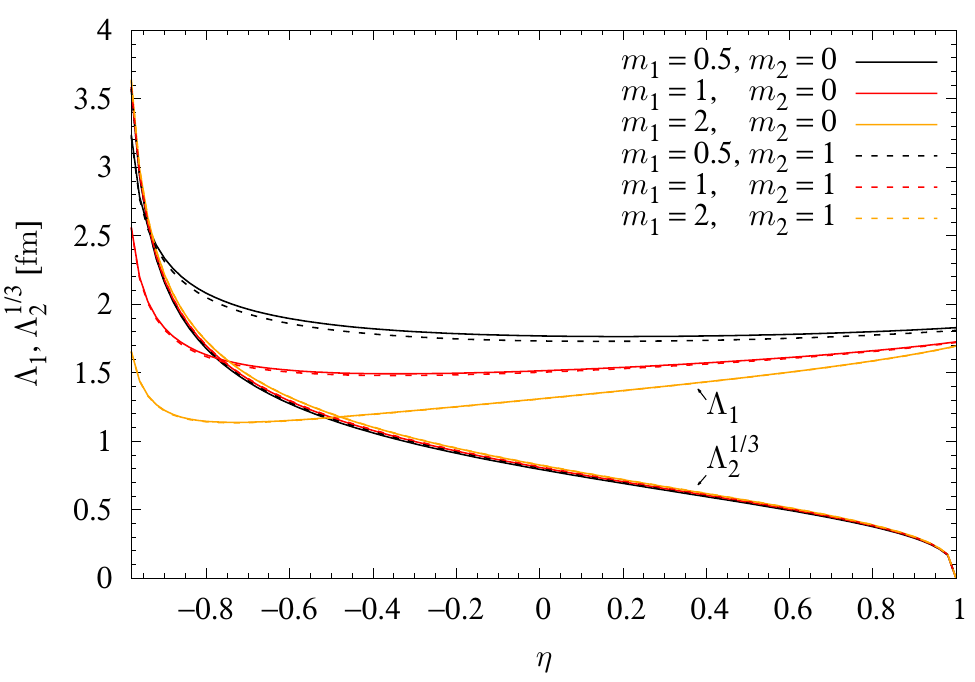}}}
  \caption{(a) The moments of inertia $\Lambda_1$, $\Lambda_2$ as
    functions of $\eta$ for a range of the pion mass parameter
    $m_1=0.5,1,2$ with and without the loosely bound potential turned
    on $m_2=0,1$.
  (b) The moments of inertia after restoring physical units:
    $\mu\lambda^2\Lambda_1$ and $(\mu\lambda^4\Lambda_2)^{1/3}$.
    }
  \label{fig:Lambdas}
\end{figure}
In Fig.~\ref{fig:Lambdas}, the moments of inertia $\Lambda_{1,2}$ are
shown.
$\Lambda_1/\Lambda_2>1$ holds for $\eta\in(-1,1]$, which is necessary
for the expansion in $1/\ell$ to be valid, but the ratio becomes
ill-defined in the limit $\eta\to-1$. 
In this limit, the stability of the Skyrmion also becomes
questionable.

\begin{figure}[!htp]
  \centering
  \mbox{\subfloat[]{\includegraphics[width=0.49\linewidth]{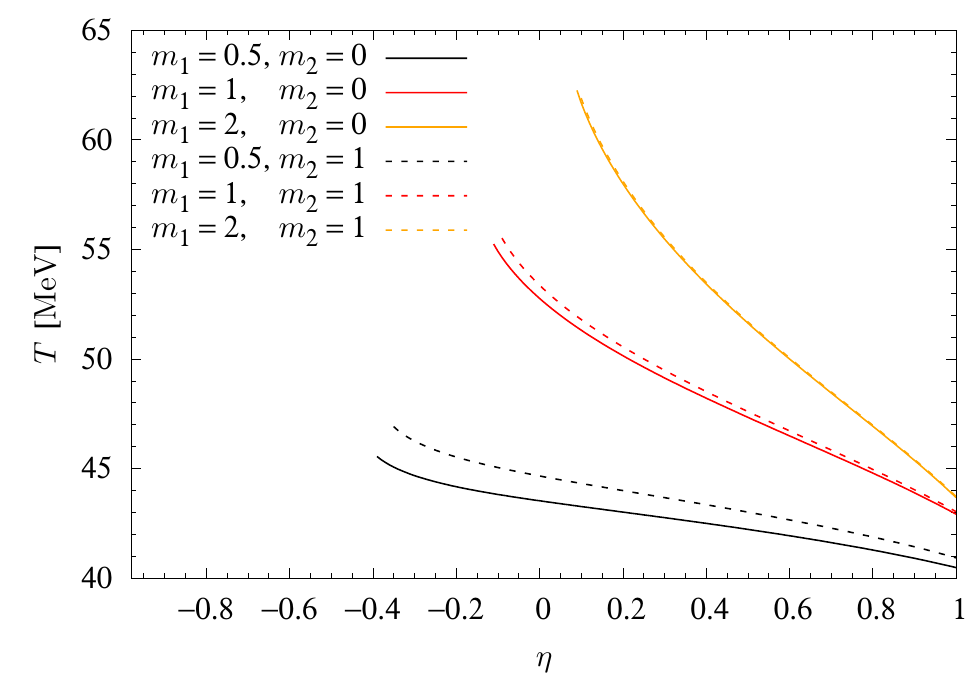}}
    \subfloat[]{\includegraphics[width=0.49\linewidth]{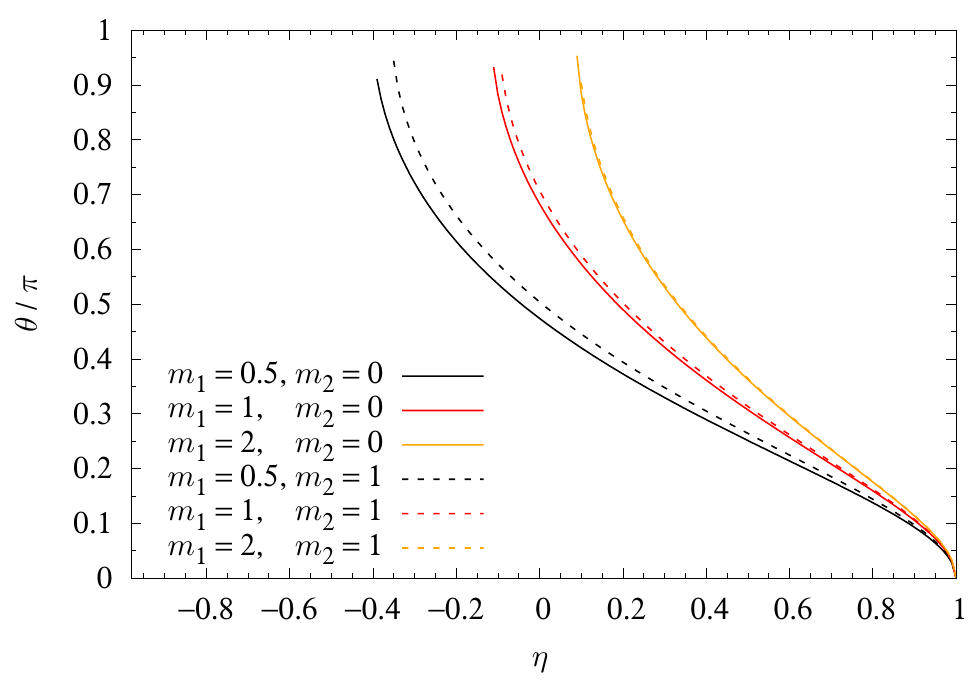}}}
  \caption{
    (a) The physical spin correction to the
    energy $T$ and (b) the angle $\theta$ of Eq.~\eqref{eq:xi_theta},
    both as functions of $\eta$ for a range of the pion mass
    parameter $m_1=0.5,1,2$ with and without the loosely bound
    potential turned on $m_2=0,1$.
    }
  \label{fig:T_theta}
\end{figure}
In Fig.~\ref{fig:T_theta}(a) we show the spin correction to
the energy as a function of $\eta$ for various pion mass and loosely
bound potential parameters, whereas in Fig.~\ref{fig:T_theta}(b) is
shown the corresponding angles $\theta$ of Eq.~\eqref{eq:xi_theta}.
We notice that the angle $\theta$ tends to $\pi$ before $\eta$ reaches
$\eta\to-1$ and hence the physical root becomes complex.
There is still a real root of Eq.~\eqref{eq:root_equation}, but it is
not connected to the physical root and it gives rise to a very large
and negative spin contribution (it is one of the unphysical roots).
This is also consistent with the results of
Ref.~\cite{Pottinger:1985fc}. 

The fit \eqref{eq:Fpi_alpha} employed here amounts to the classical
mass of the Skyrmion always being at the experimental face value and
so the spin correction should be as small as possible.
Re-calibration could of course get the nucleon mass right, but as
discussed in the introduction, the larger the spin energy is, the
larger the binding energies are.
Since they should physically be around $8\MeV$ per nucleon for larger
nuclei, a spin correction to the energy much larger than that creates
tension and warrants other (extended) quantization methods to provide
physical spectra, see e.g.~Ref.~\cite{Gudnason:2023jpq}.
We note that the dependence of the spin correction on the pion mass
parameter $m_1$ is quite large, whereas the dependence on the loosely
bound potential parameter $m_2$ is only mild.

\begin{figure}[!htp]
  \centering
  \mbox{\subfloat[]{\includegraphics[width=0.49\linewidth]{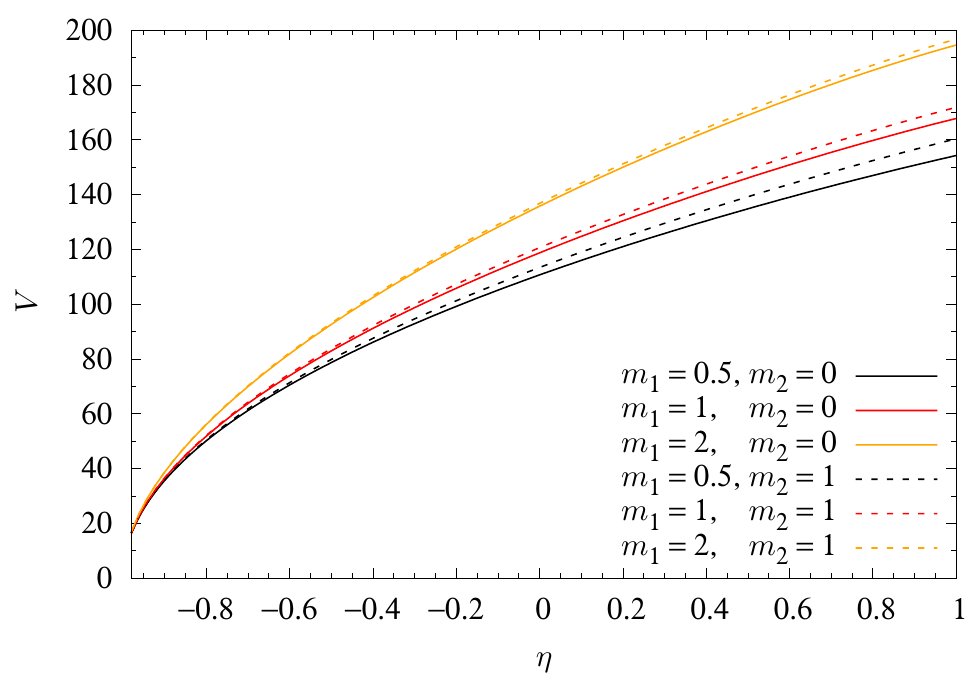}}
    \subfloat[]{\includegraphics[width=0.49\linewidth]{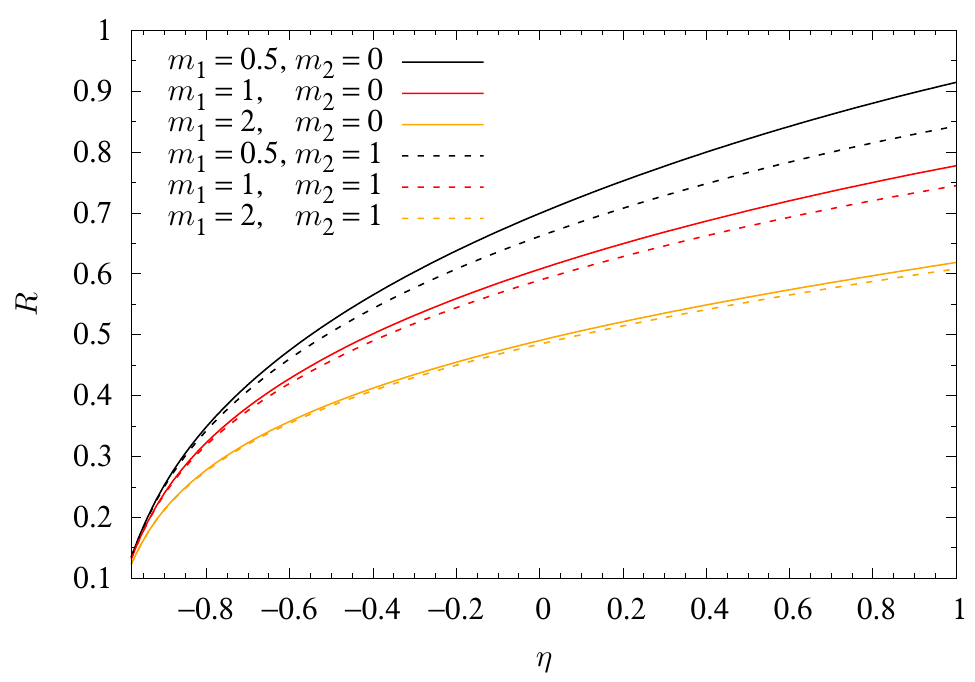}}}
  \caption{
    (a) The static energy of the Skyrmion in Skyrme units and (b) the (charge)
    radius of the Skyrmion, both as functions of $\eta$ for a range of
    the pion mass parameter $m_1=0.5,1,2$ with and without the loosely
    bound potential turned on $m_2=0,1$.
  }
  \label{fig:V_R}
\end{figure}
Fig.~\ref{fig:V_R} shows the static classical mass and radius of the
Skyrmion in dimensionless (model) units.
We plot the figures only in the range $\eta\in[-0.98,1]$, since we are
unable to obtain trustable solutions in the limit $\eta\to-1$, where
the Skyrmion size is also seen to shrink to zero
(Fig.~\ref{fig:V_R}(b)), which is also anticipated in
appendices \ref{app:positivity_static_PR} and \ref{app:topobound_PR}.
In the latter appendix, we also note that the topological energy bound
goes to zero in the $\eta\to-1$ limit. 

In physical units, the classical mass and radius are exactly equal to
their experimental values, $939\MeV$ and $0.8783\fm$, respectively,
due to the calibration of the model \eqref{eq:Fpi_alpha}.

\begin{figure}[!htp]
  \centering
  \mbox{\subfloat[]{\includegraphics[width=0.49\linewidth]{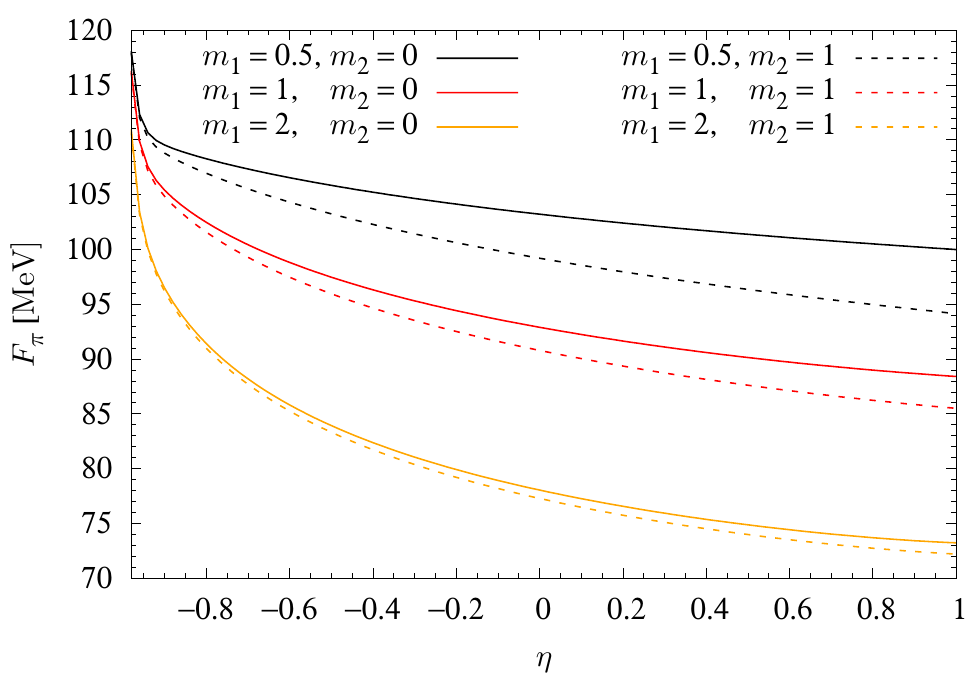}}
    \subfloat[]{\includegraphics[width=0.49\linewidth]{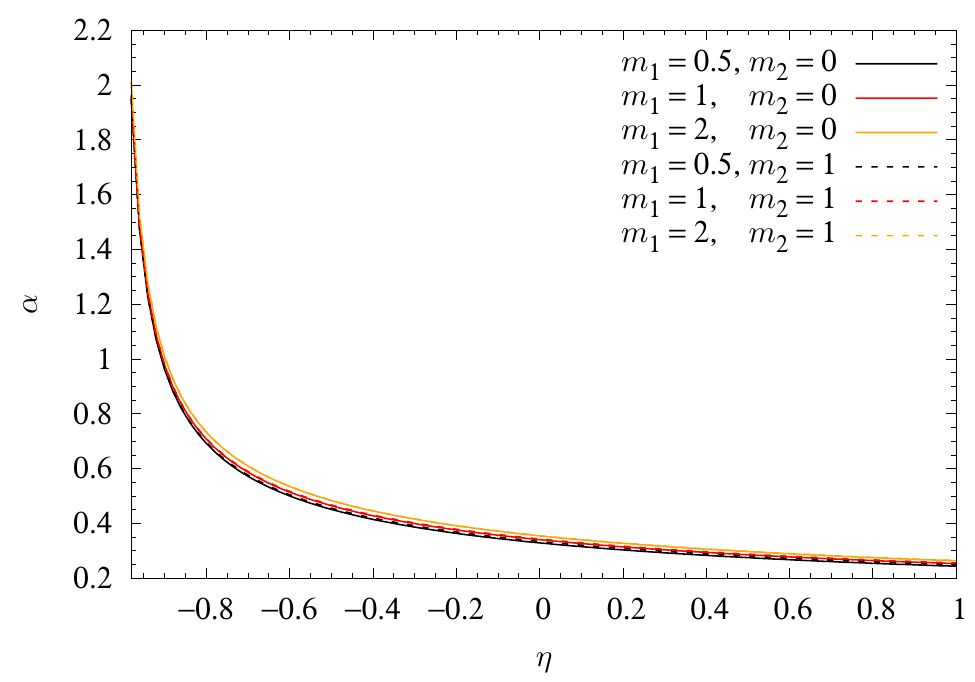}}}
  \caption{(a) The pion decay constant and (b) the four-derivative
    coupling constant $\alpha$, both as functions of $\eta$
    for a range of the pion mass parameter
    $m_1=0.5,1,2$ with and without the loosely bound potential turned
    on $m_2=0,1$.
    The physical value of the pion decay constant in the
    normalization used in this paper is $186\MeV$.
  }
  \label{fig:Fpi_alpha}
\end{figure}
\begin{figure}[!htp]
  \centering
  \mbox{\subfloat[]{\includegraphics[width=0.49\linewidth]{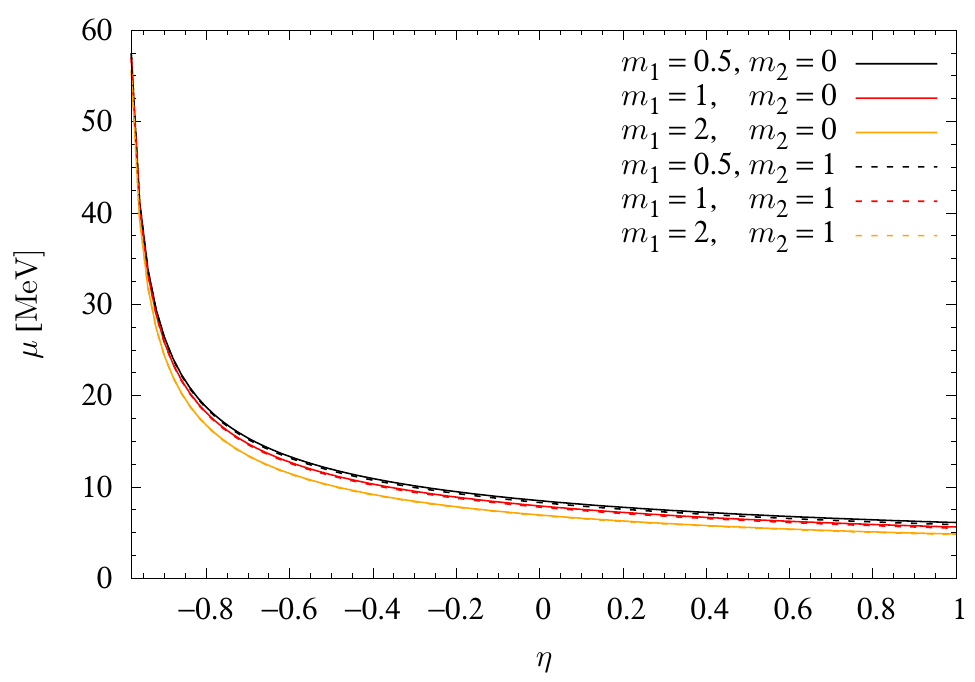}}
    \subfloat[]{\includegraphics[width=0.49\linewidth]{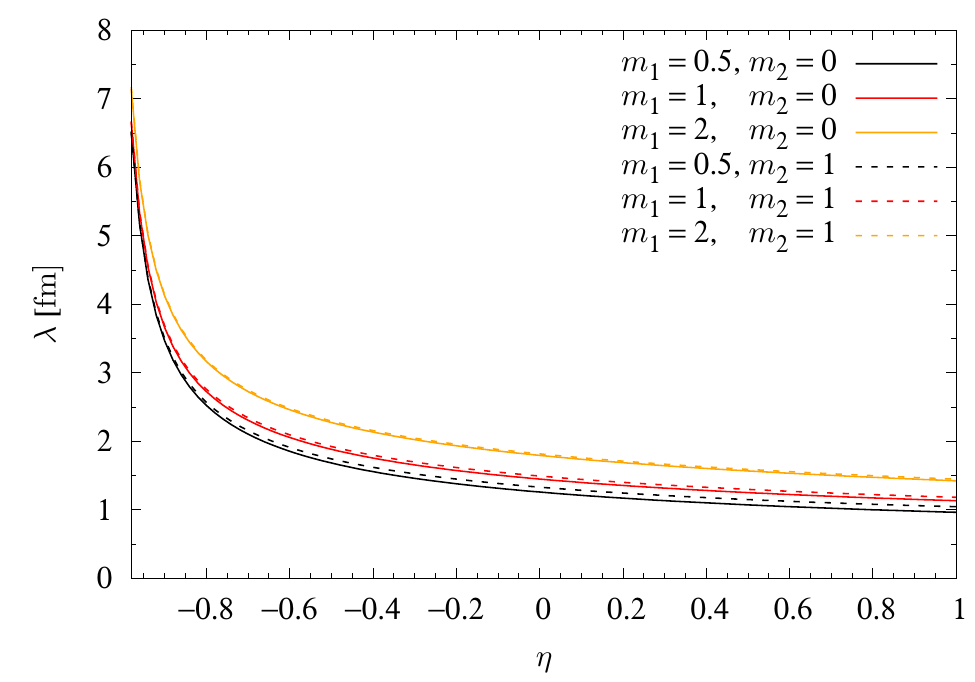}}}
  \caption{The calibrated (a) energy scale and (b) length scale of the
  model, both as functions of $\eta$
    for a range of the pion mass parameter
    $m_1=0.5,1,2$ with and without the loosely bound potential turned
    on $m_2=0,1$.
  }
  \label{fig:mu_lambda}
\end{figure}
\begin{figure}[!htp]
  \centering
  \includegraphics[width=0.49\linewidth]{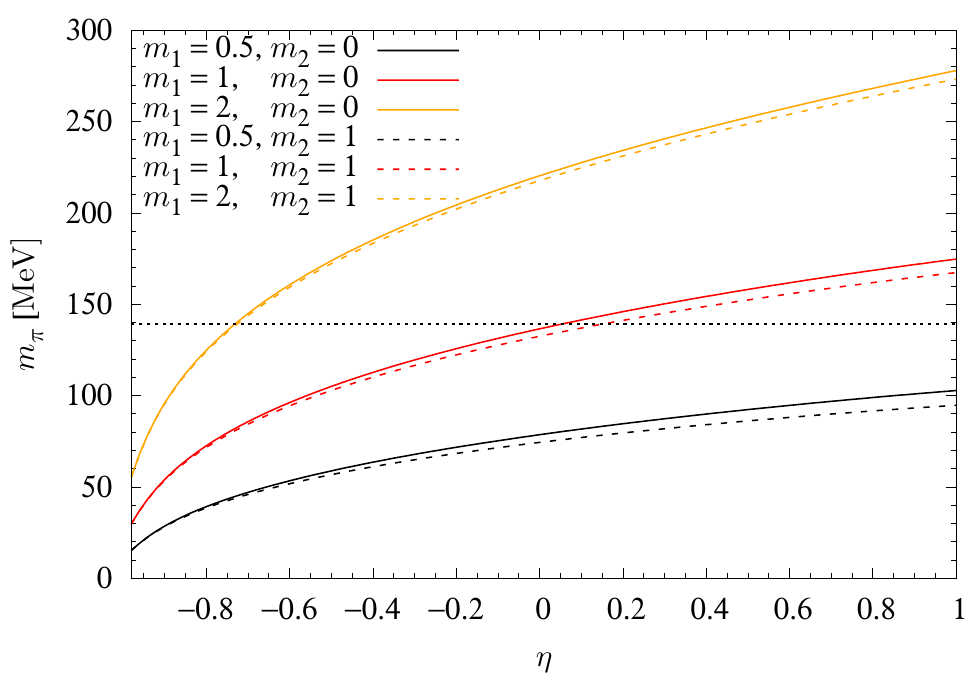}
  \caption{The pion mass as a function of $\eta$
    for a range of the pion mass parameter
    $m_1=0.5,1,2$ with and without the loosely bound potential turned
    on $m_2=0,1$.
    The physical value of the pion mass is around $139\MeV$ (ignoring
    isospin breaking effects) and is shown with a dotted horizontal
    black line.
  }
  \label{fig:mpi}
\end{figure}
Having the classical mass and radius of the Skyrmion in hand, we thus
readily calibrate the model according to Eq.~\eqref{eq:Fpi_alpha},
yielding the pion decay constant in Fig.~\ref{fig:Fpi_alpha}(a), the
four-derivative term coupling constant $\alpha$ in
Fig.~\ref{fig:Fpi_alpha}(b) or equivalently the energy scale $\mu$ in
Fig.~\ref{fig:mu_lambda}(a) and the length scale $\lambda$ in
Fig.~\ref{fig:mu_lambda}(b).
For completeness, we plot the pion mass in Fig.~\ref{fig:mpi}, from
which we can see that the pion mass parameter should be taken
somewhere between $0.75$ and $1$, depending on the values of $\eta$ 
in its physical regime (see Fig.~\ref{fig:T_theta}) and $m_2$, 
in order to reproduce the experimental value of roughly $139\MeV$.

\section{Higher-order Skyrme models}\label{sec:higherorder}

We now consider the cases of the higher-order models introduced in
Ref.~\cite{Gudnason:2017opo}, i.e.~higher-order derivative theories
with four time derivatives and with eight to twelve derivatives in
total.
The Lagrangian reads
\begin{align}
  \Lag &= \frac{F_\pi^2}{16}\tr(R_\mu R^\mu)
  +\frac{1}{32e^2}\tr\left([R_\mu,R_\nu][R^\mu,R^\nu]\right)
  +\Lag'\non
  &\phantom{=\ }
  -\frac{F_\pi^2m_\pi^2}{8}\tr(\mathbf{1}_2 - U )
  -\frac{F_\pi^2M^2}{32}\left[\tr(\mathbf{1}_2 - U)\right]^2,
  \label{eq:LHOaphysicalunits}
\end{align}
with the new higher-derivative term $\Lag'$ being one of the four
possibilities:
\begin{align}
  \Lag_{8a} &=
  -\frac{\beta_8}{1024F_\pi^4}\left(\tr\left([R_\mu,R_\nu][R^\mu,R^\nu]\right)\right)^2,\\
  \Lag_{8b} &=
  -\frac{\beta_8}{768F_\pi^4}\tr(R_\sigma R^\sigma)\tr\left([R_\mu,R^\nu][R_\nu,R^\rho][R_\rho,R^\mu]\right),\\
  \Lag_{10} &=
  -\frac{\beta_{10}}{3072F_\pi^6}\tr\left([R_\sigma,R_\delta][R^\sigma,R^\delta]\right)\tr\left([R_\mu,R^\nu][R_\nu,R^\rho][R_\rho,R^\mu]\right),\\
  \Lag_{12} &=
  -\frac{\beta_{12}}{9216F_\pi^8}\left(\tr\left([R_\mu,R^\nu][R_\nu,R^\rho][R_\rho,R^\mu]\right)\right)^2.
\end{align}
We first define the relations to the coupling $\eta$:
\beq
\frac{1}{e^2} = \alpha^2\eta, \qquad
\beta_n = \big(\sqrt2\alpha\big)^{n-2}(1-\eta),
\eeq
with $\eta\in[0,1]$.
For the analysis of the positivity of the static energy leading to the
viable range of the parameter $\eta$, see
App.~\ref{app:positivity_static_HO}.
Rescaling the Lagrangians to the energy and length units
\eqref{eq:mu_lambda} yields the dimensionless Lagrangian
\begin{align}
  \Lag &= \frac12\tr(R_\mu R^\mu)
  +\frac{\eta}{16}\tr\left([R_\mu,R_\nu][R^\mu,R^\nu]\right)
  +\Lag'\non
  &\phantom{=\ }
  -m_1^2\tr(\mathbf{1}_2 - U )
  -\frac{m_2^2}{4}\left[\tr(\mathbf{1}_2 - U)\right]^2,
  \label{eq:LHO}
\end{align}
with the dimensionless higher-order term $\Lag'$ being one of the
following four terms:
\begin{align}
  \Lag_{8a} &=
  -\frac{1-\eta}{1024}\left(\tr\left([R_\mu,R_\nu][R^\mu,R^\nu]\right)\right)^2,\label{eq:L8a}\\
  \Lag_{8b} &=
  -\frac{1-\eta}{768}\tr(R_\sigma R^\sigma)\tr\left([R_\mu,R^\nu][R_\nu,R^\rho][R_\rho,R^\mu]\right),\label{eq:L8b}\\
  \Lag_{10} &=
  -\frac{1-\eta}{6144}\tr\left([R_\sigma,R_\delta][R^\sigma,R^\delta]\right)\tr\left([R_\mu,R^\nu][R_\nu,R^\rho][R_\rho,R^\mu]\right),\label{eq:L10}\\
  \Lag_{12} &=
  -\frac{1-\eta}{36864}\left(\tr\left([R_\mu,R^\nu][R_\nu,R^\rho][R_\rho,R^\mu]\right)\right)^2.\label{eq:L12}
\end{align}
Splitting the Lagrangians up into potential and kinetic terms, we get
\begin{align}
  L &= T^L - V, \\
  V &= \int\d^3x\bigg[
    -\frac12\tr(R_i^2)
    -\frac{\eta}{16}\tr\left([R_i,R_j]^2\right)
    +m_1^2\tr(\mathbf{1}_2 - U)
    +\frac{m_2^2}{4}\left[\tr(\mathbf{1}_2 - U)\right]^2
  \bigg] + V',\label{eq:VHO}\\
  T^L &= \int\d^3x\bigg[
    -\frac12\tr(T_iT_j)
    -\frac{\eta}{8}\tr\left([T_i,R_k][T_j,R_k]\right)
  \bigg]a_i a_j + {T^L}',
\end{align}
with potential terms
\begin{align}
  V_{8a} &= \frac{1-\eta}{1024}\int\d^3x\left(\tr\left([R_i,R_j]^2\right)\right)^2,\label{eq:VHO_8a}\\
  V_{8b} &= \frac{1-\eta}{768}\int\d^3x\tr(R_l^2)\tr\left([R_i,R_j][R_j,R_k][R_k,R_i]\right),\label{eq:VHO_8b}\\
  V_{10} &= \frac{1-\eta}{6144}\int\d^3x\tr\left([R_k,R_l]^2\right)\tr\left([R_i,R_j][R_j,R_k][R_k,R_i]\right),\label{eq:VHO_10}\\
  V_{12} &= \frac{1-\eta}{36864}\int\d^3x\;\big(\tr\left([R_i,R_j][R_j,R_k][R_k,R_i]\right)\big)^2,\label{eq:VHO_12}
\end{align}
and kinetic terms
\begin{align}
  T_{8a}^{L} &= \frac{1-\eta}{256}\int\d^3x\;\tr\left([T_i,R_k][T_j,R_k]\right)\tr\left([R_l,R_m]^2\right)a_i a_j\non
  &\phantom{=\ }
  -\frac{1-\eta}{256}\int\d^3x\;\tr\left([T_i,R_m][T_j,R_m]\right)\tr\left([T_k,R_n][T_l,R_n]\right)a_i a_j a_k a_l,\\
  T_{8b}^{L} &= \frac{1-\eta}{768}\int\d^3x\Big[
    \tr(T_iT_j)\tr\left([R_k,R_l][R_l,R_m][R_m,R_k]\right)\non
    &\phantom{=\frac{1-\eta}{1536}\int\d^3x\Big[\ }
    +3\tr(R_m^2)\tr\left([T_i,R_k][R_k,R_l][R_l,T_j]\right)\Big]a_i a_j\non
  &\phantom{=\ }
    -\frac{1-\eta}{256}\int\d^3x\;\tr(T_iT_j)\tr\left([T_k,R_m][R_m,R_n][R_n,T_l]\right)a_i a_j a_k a_l,\\
  T_{10}^{L} &= \frac{1-\eta}{6144}\int\d^3x\Big[
    2\tr\left([T_i,R_k][T_j,R_k]\right)\tr\left([R_l,R_m][R_m,R_n][R_n,R_l]\right)\non
    &\phantom{=\frac{1-\eta}{6144}\int\d^3x\Big[\ }
    +3\tr\left([R_m,R_n]^2\right)\tr\left([T_i,R_k][R_k,R_l][R_l,T_j]\right)\Big]a_i a_j\non
  &\phantom{=\ }
    -\frac{1-\eta}{1024}\int\d^3x\;\tr\left([T_i,R_o][T_j,R_o]\right)\tr\left([T_k,R_m][R_m,R_n][R_n,T_l]\right)a_i a_j a_k a_l,\\
  T_{12}^{L} &= \frac{1-\eta}{6144}\int\d^3x\;
  \tr\left([T_i,R_k][R_k,R_l][R_l,T_j]\right)\tr\left([R_m,R_n][R_o,R_p][R_p,R_m]\right)a_i a_j\non
  &\phantom{=\ }
  -\frac{1-\eta}{4096}\int\d^3x\;
  \tr\left([T_i,R_m][R_m,R_n][R_n,T_j]\right)\tr\left([T_k,R_o][R_o,R_p][R_p,T_l]\right)a_i a_j a_k a_l,
\end{align}
and $T_i=\frac{\i}{2}[\tau^i,U]U^\dag$ as always.
Inserting the hedgehog Ansatz \eqref{eq:hedgehog}, we obtain the
potential term
\begin{align}
  V &= \int\d^3x\bigg[
    (f')^2 + \frac{2\sin^2f}{r^2}
    +\eta\frac{\sin^2f}{r^4}\left(\sin^2f+2r^2(f')^2\right)\non
    &\phantom{=\int\d^3x\bigg[\ }
    +2m_1^2(1-\cos f)
    +m_2^2(1-\cos f)^2
    \bigg] + V'
\end{align}
with
\begin{align}
  V_{8a} &= \frac{1-\eta}{4}\int\d^3x\;\frac{\sin^4f}{r^8}\left(\sin^2f+2r^2(f')^2\right)^2,\\
  V_{8b} &= \frac{1-\eta}{4}\int\d^3x\;\frac{\sin^4(f)(f')^2}{r^6}\left(2\sin^2f+r^2(f')^2\right),\\
  V_{10} &= \frac{1-\eta}{4}\int\d^3x\;\frac{\sin^6(f)(f')^2}{r^8}\left(\sin^2f+2r^2(f')^2\right),\\
  V_{12} &= \frac{1-\eta}{4}\int\d^3x\;\frac{\sin^8(f)(f')^4}{r^8},
\end{align}
whereas for the kinetic term we get
\beq
T^L = \frac12\Lambda_1a_i^2 - \frac14\Lambda_2'a_i^2a_j^2,
\eeq
with
\beq
\Lambda_1 = \frac{16\pi}{3}\int\d r\; r^2\bigg[
    \sin^2f
    +\eta\sin^2f\left(\frac{\sin^2f}{r^2} + (f')^2\right)
    \bigg] + \Lambda_1',
\eeq
and
\begin{align}
  \Lambda_1^{(8a)} &= \frac{8\pi(1-\eta)}{3}\int\d r\;
  \frac{\sin^4f}{r^4}\left(\sin^2f+r^2(f')^2\right)\left(\sin^2f+2r^2(f')^2\right),\\
  \Lambda_1^{(8b)} &= \frac{4\pi(1-\eta)}{3}\int\d r\;
  \frac{\sin^4(f)(f')^2}{r^2}\left(3\sin^2f+r^2(f')^2\right),\\
  \Lambda_1^{(10)} &= \frac{4\pi(1-\eta)}{3}\int\d r\;
  \frac{\sin^6(f)(f')^2}{r^4}\left(2\sin^2f+3r^2(f')^2\right),\\
  \Lambda_1^{(10)} &= \frac{8\pi(1-\eta)}{3}\int\d r\;
  \frac{\sin^8(f)(f')^4}{r^4},
\end{align}
as well as
\begin{align}
  \Lambda_2^{(8a)} &= \frac{32\pi(1-\eta)}{15}\int\d r\;
  \frac{\sin^4f}{r^2}\left(\sin^2f+r^2(f')^2\right)^2,\\
  \Lambda_2^{(8b)} &= \frac{32\pi(1-\eta)}{15}\int\d r\;\sin^6(f)(f')^2,\\
  \Lambda_2^{(10)} &= \frac{32\pi(1-\eta)}{15}\int\d r\;
  \frac{\sin^6(f)(f')^2}{r^2}\left(\sin^2f+r^2(f')^2\right),\\
  \Lambda_2^{(12)} &= \frac{32\pi(1-\eta)}{15}\int\d r\;
  \frac{\sin^8(f)(f')^4}{r^2}.
\end{align}
The quantum kinetic energy is then given by Eq.~\eqref{eq:T} and using
the $\Lambda_{1,2}$ of the higher-order model of this section, we can
readily use the result for the spin correction to the energy
\eqref{eq:Tphys_units}.
This result follows through because both the PR model and the
higher-order models have exactly 4 time derivatives.
We will again calibrate the model in the same way using
Eq.~\eqref{eq:Fpi_alpha}, but with the energy given by the potential
\eqref{eq:VHO} and the size computed from the solutions to the
higher-order model, see Sec.~\ref{sec:calib}.

\begin{figure}[!tp]
  \centering
  \mbox{\subfloat[]{\includegraphics[width=0.49\linewidth]{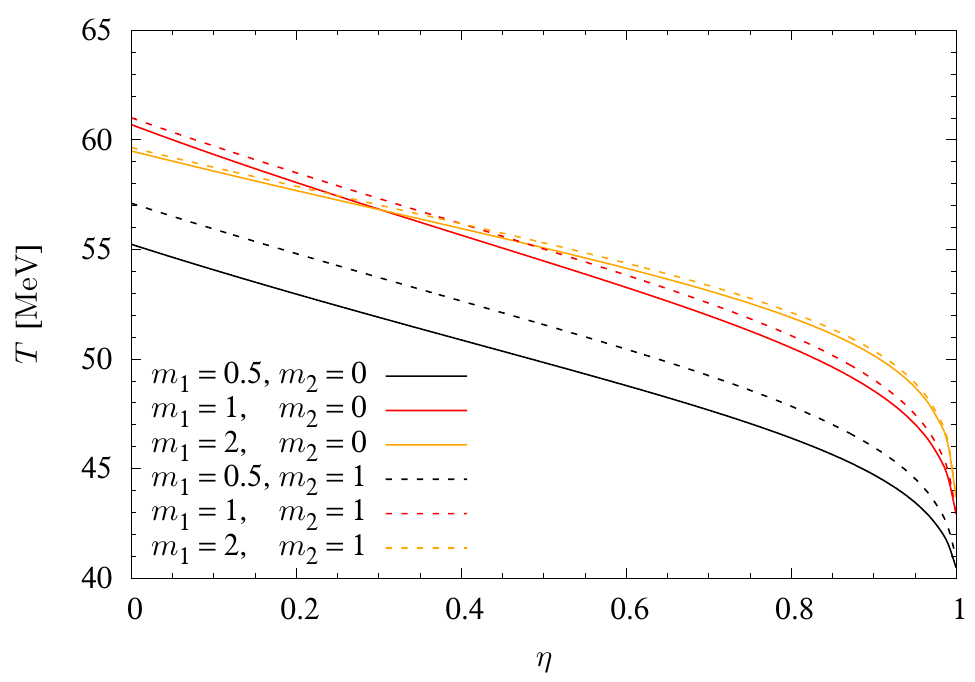}}
    \subfloat[]{\includegraphics[width=0.49\linewidth]{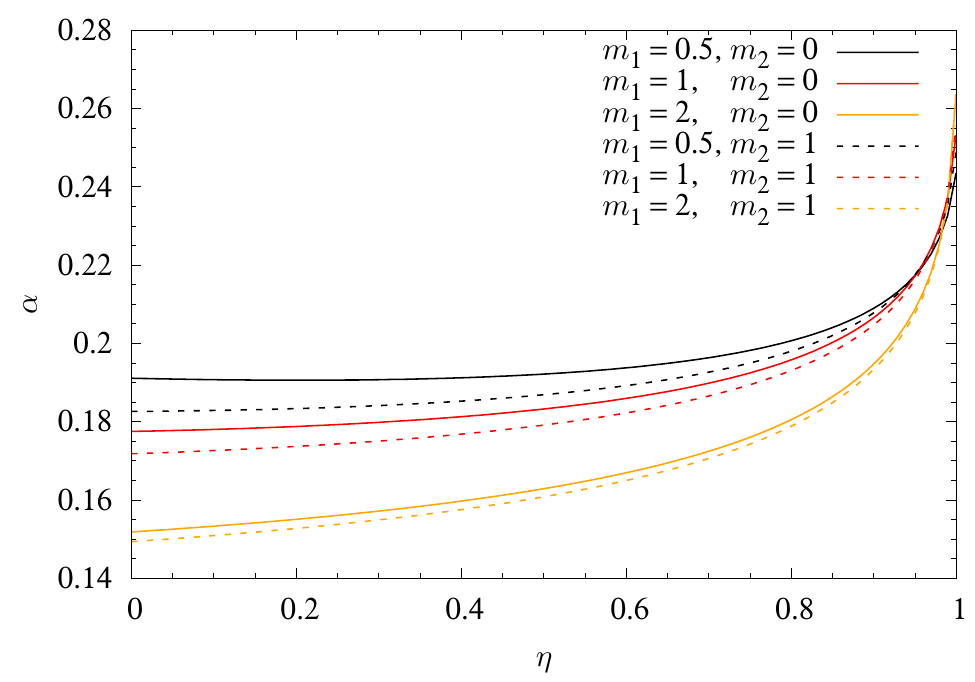}}}
  \mbox{\subfloat[]{\includegraphics[width=0.49\linewidth]{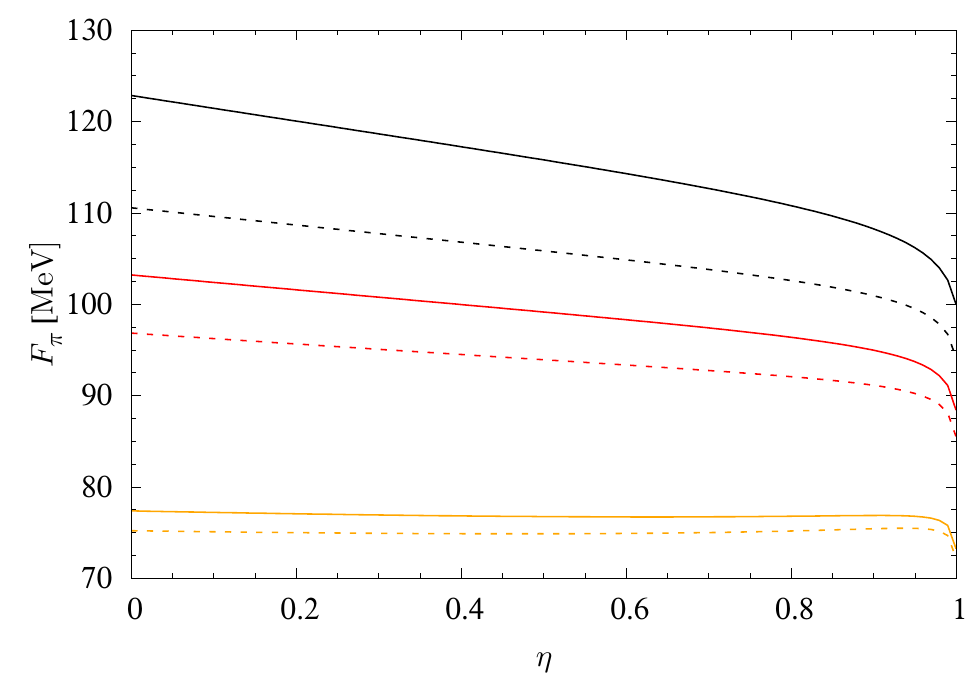}}
    \subfloat[]{\includegraphics[width=0.49\linewidth]{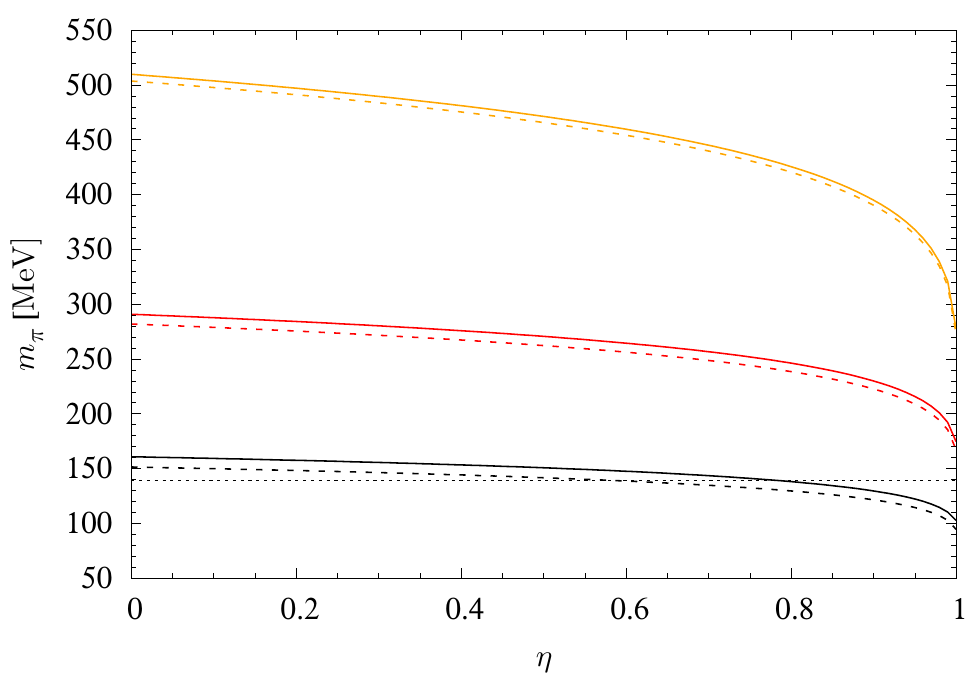}}}
  \caption{The (a) spin correction to the energy, (b) the dimensional
    coupling constant of the Skyrme and higher-order terms, (c) the
    pion decay constant and (d) the pion mass, all as functions of
    $\eta\in[0,1]$ which interpolates between the Skyrme model
    ($\eta=1$) and the higher-order model ($\eta=0$) $\Lag_{8a}$. }
  \label{fig:HO8a}
\end{figure}

\begin{figure}[!tp]
  \centering
  \mbox{\subfloat[]{\includegraphics[width=0.49\linewidth]{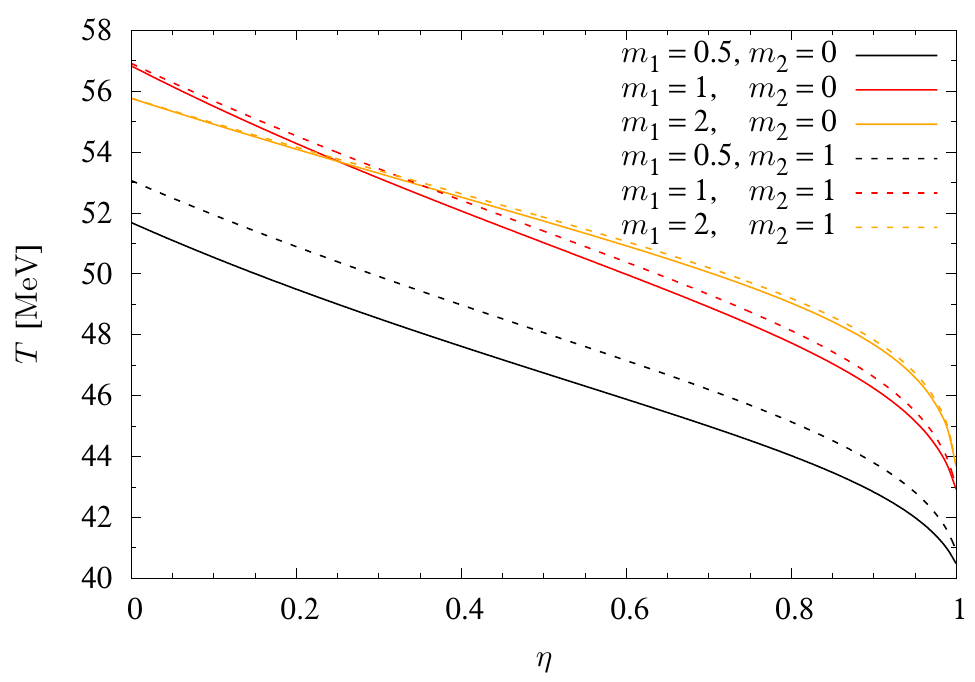}}
    \subfloat[]{\includegraphics[width=0.49\linewidth]{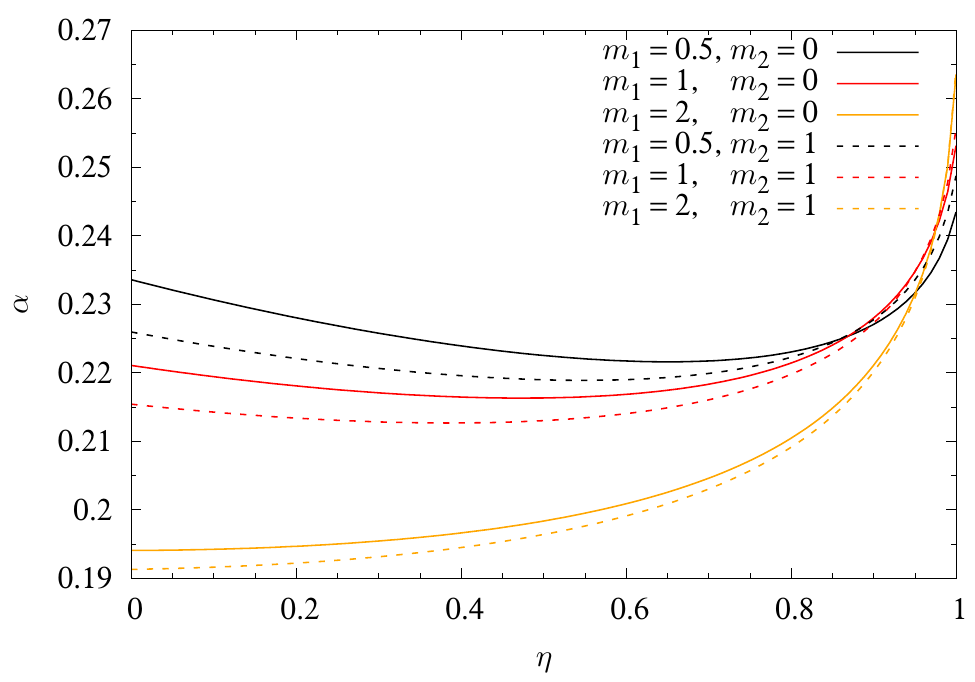}}}
  \mbox{\subfloat[]{\includegraphics[width=0.49\linewidth]{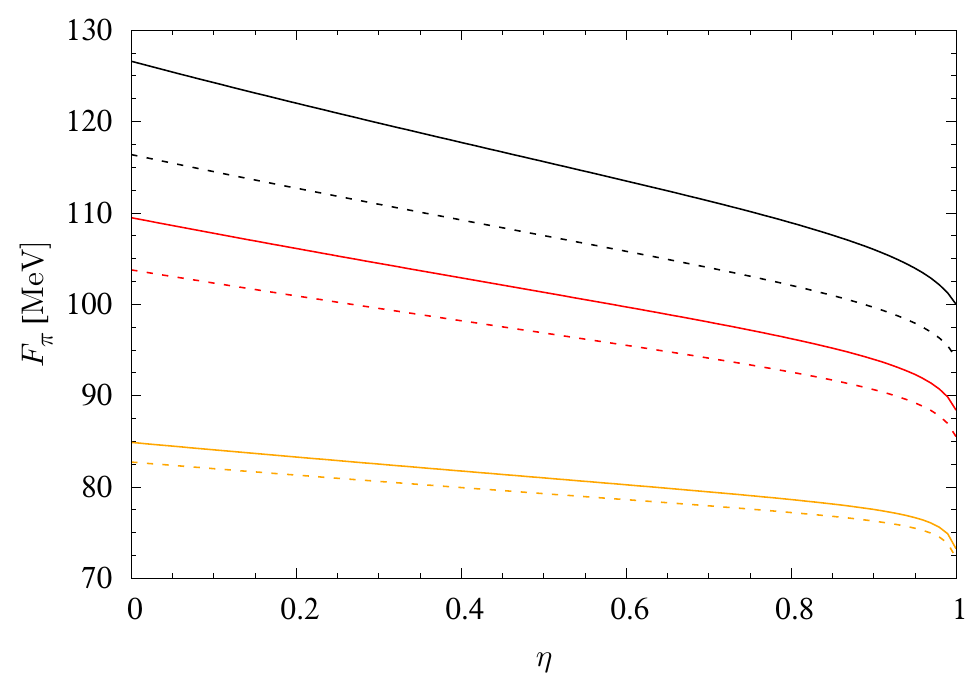}}
    \subfloat[]{\includegraphics[width=0.49\linewidth]{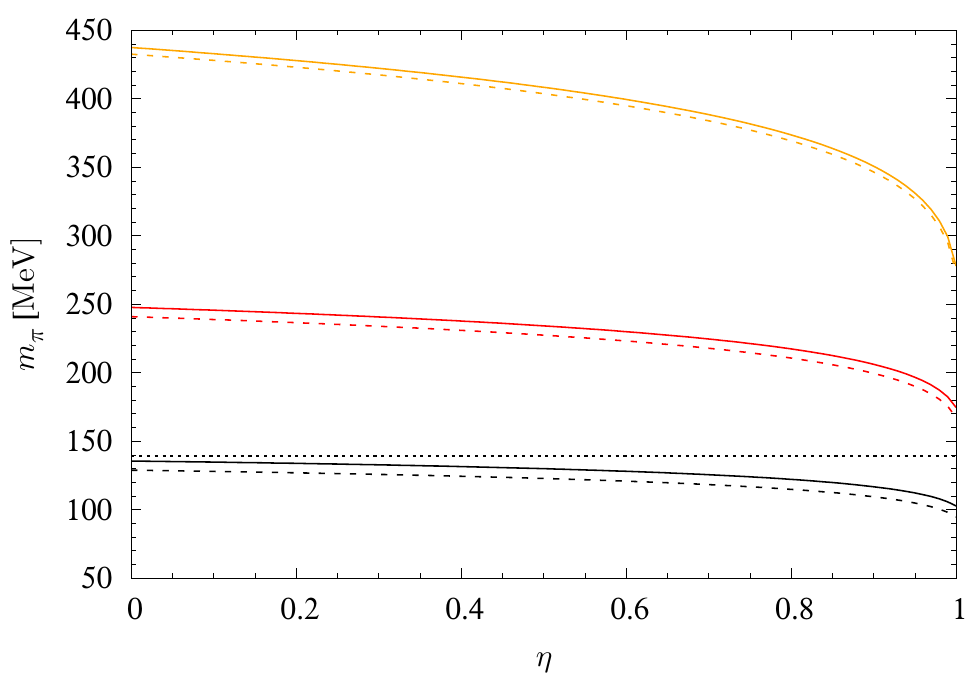}}}
  \caption{The (a) spin correction to the energy, (b) the dimensional
    coupling constant of the Skyrme and higher-order terms, (c) the
    pion decay constant and (d) the pion mass, all as functions of
    $\eta\in[0,1]$ which interpolates between the Skyrme model
    ($\eta=1$) and the higher-order model ($\eta=0$) $\Lag_{8b}$. }
  \label{fig:HO8b}
\end{figure}

\begin{figure}[!tp]
  \centering
  \mbox{\subfloat[]{\includegraphics[width=0.49\linewidth]{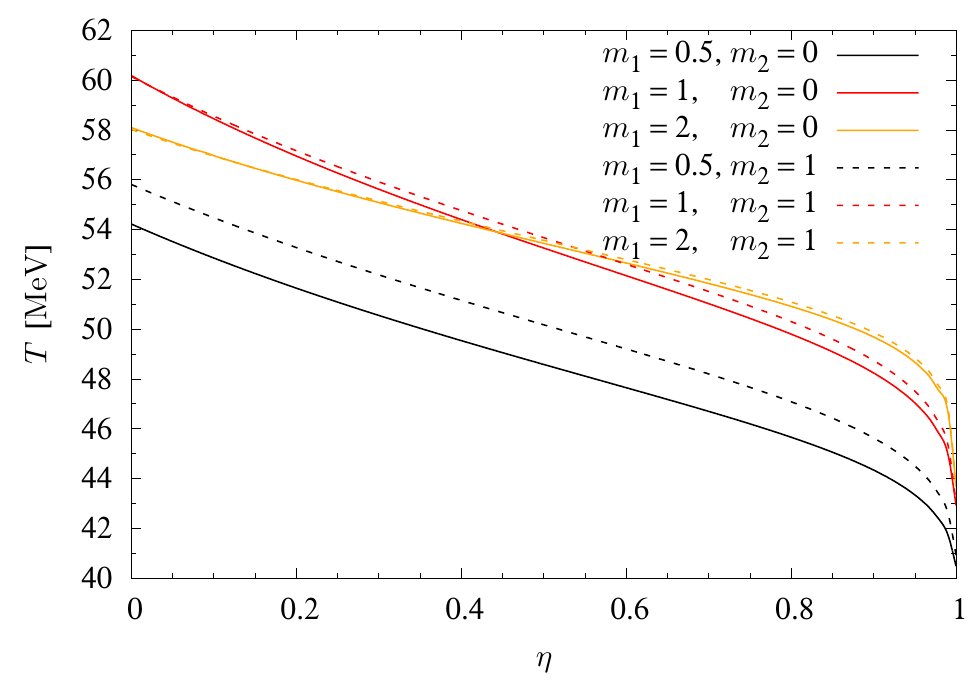}}
    \subfloat[]{\includegraphics[width=0.49\linewidth]{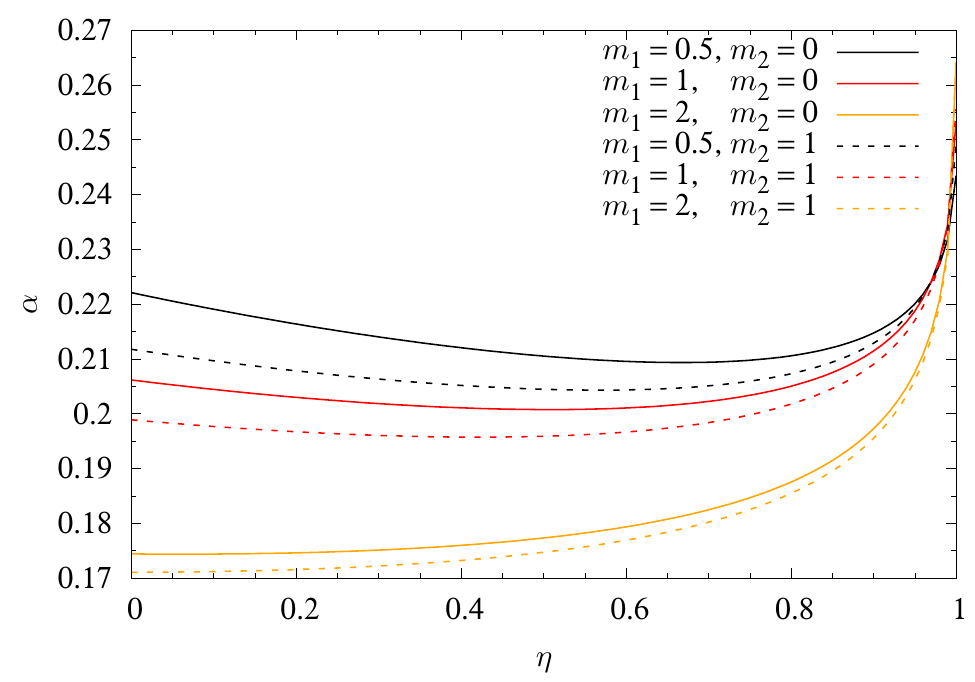}}}
  \mbox{\subfloat[]{\includegraphics[width=0.49\linewidth]{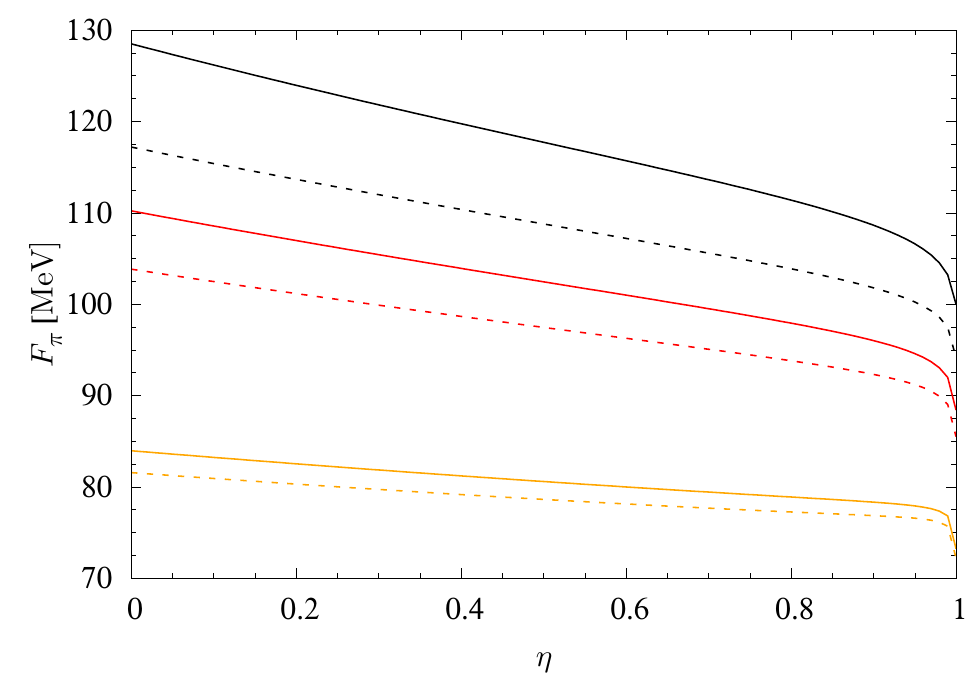}}
    \subfloat[]{\includegraphics[width=0.49\linewidth]{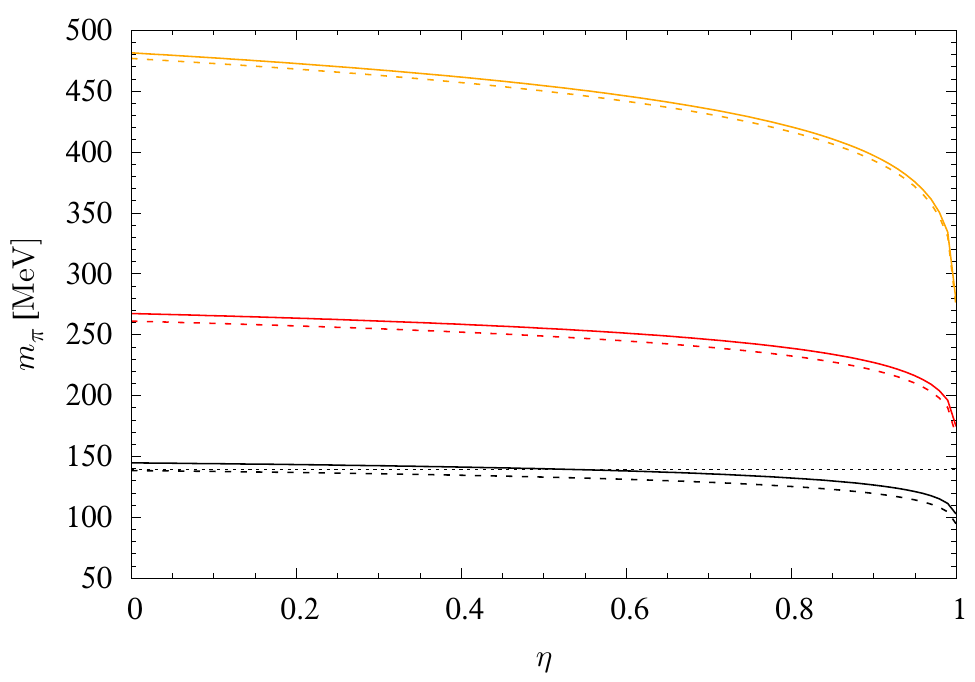}}}
  \caption{The (a) spin correction to the energy, (b) the dimensional
    coupling constant of the Skyrme and higher-order terms, (c) the
    pion decay constant and (d) the pion mass, all as functions of
    $\eta\in[0,1]$ which interpolates between the Skyrme model
    ($\eta=1$) and the higher-order model ($\eta=0$) $\Lag_{10}$. }
  \label{fig:HO10}
\end{figure}

\begin{figure}[!tp]
  \centering
  \mbox{\subfloat[]{\includegraphics[width=0.49\linewidth]{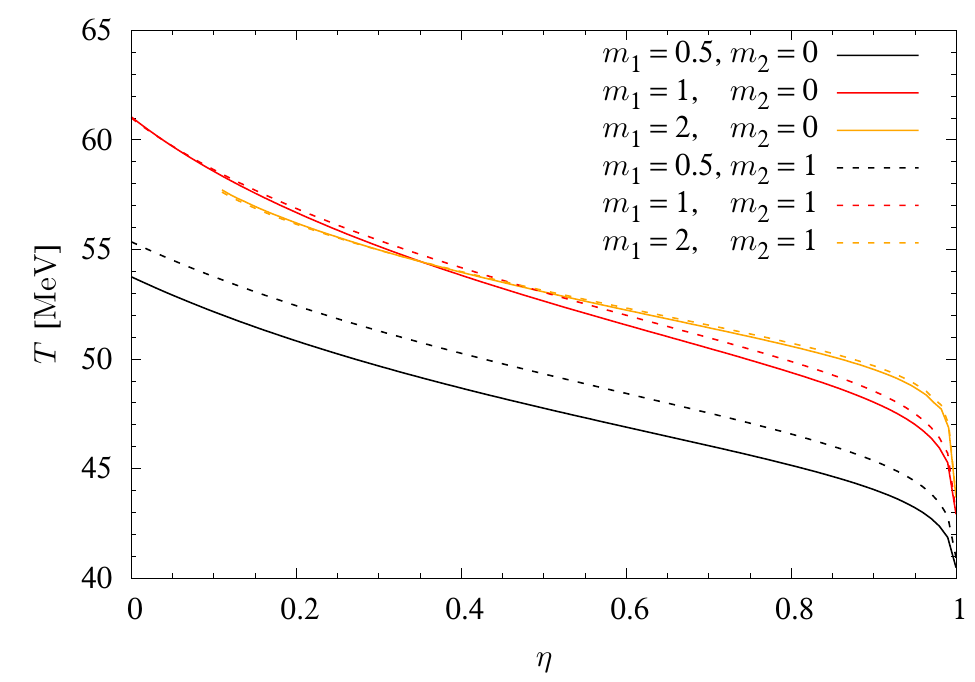}}
    \subfloat[]{\includegraphics[width=0.49\linewidth]{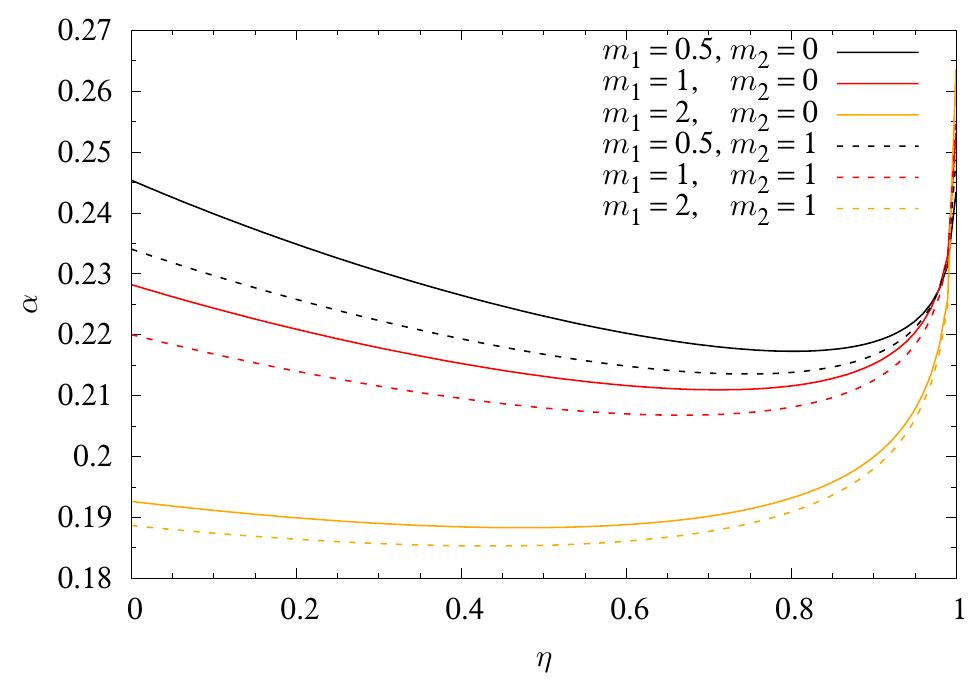}}}
  \mbox{\subfloat[]{\includegraphics[width=0.49\linewidth]{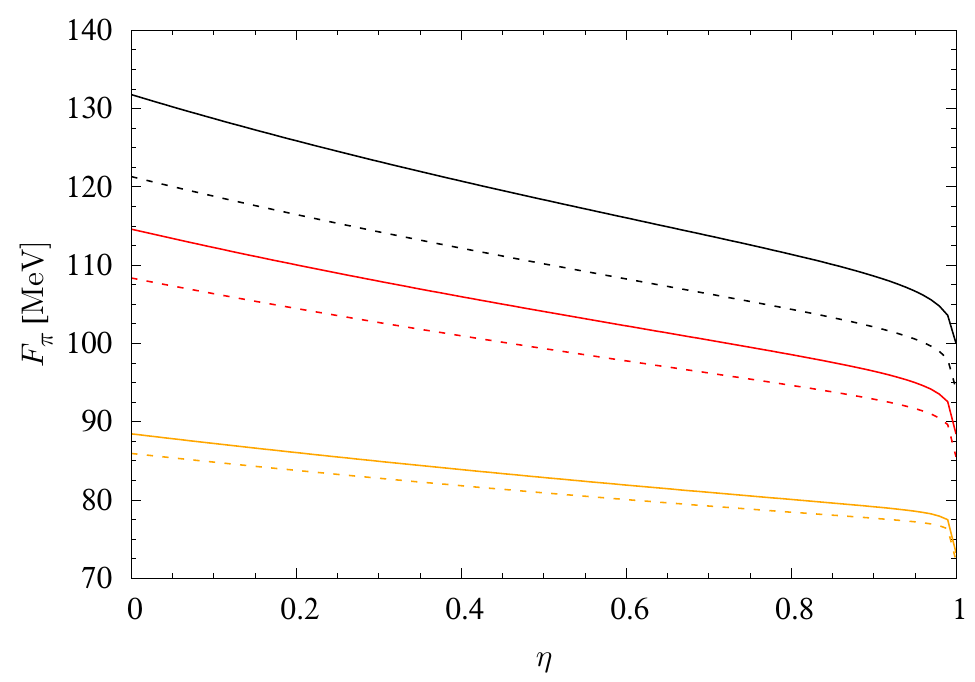}}
    \subfloat[]{\includegraphics[width=0.49\linewidth]{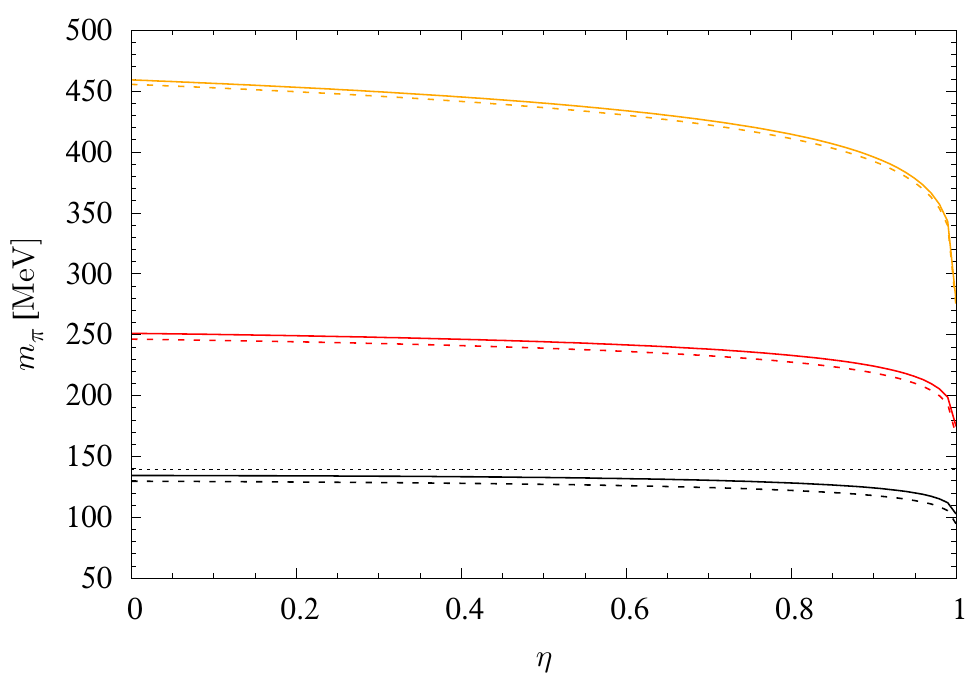}}}
  \caption{The (a) spin correction to the energy, (b) the dimensional
    coupling constant of the Skyrme and higher-order terms, (c) the
    pion decay constant and (d) the pion mass, all as functions of
    $\eta\in[0,1]$ which interpolates between the Skyrme model
    ($\eta=1$) and the higher-order model ($\eta=0$) $\Lag_{12}$. }
  \label{fig:HO12}
\end{figure}

We are now ready to present the numerical results for the higher-order
Skyrme models \eqref{eq:L8a}-\eqref{eq:L12}, which are shown in
Figs.~\ref{fig:HO8a}-\ref{fig:HO12}.
In the figures, the spin corrections to the energy is shown in panels
(a), the dimensionless coupling constant $\alpha$ in panels (b), the
pion decay constant in panels (c) and finally, the pion mass in panels
(d).
It is seen also for all the higher-order models, that the spin
correction to the energy is smallest in the Skyrme model limit,
i.e.~for $\eta=1$.
We also notice that all the models have viable (calculable) spin
contributions to the energy in the entire parameter space,
i.e.~$\eta\in[0,1]$, except in the case of the 12th-order model
\eqref{eq:L12}, where the spin contribution to the energy disappears
for $\eta\lesssim0.12$ for $m_1=2$ (very large pion mass), since the
root in Eq.~\eqref{eq:root_equation} turns complex and hence
unphysical.
This should not be too worrisome, since $m_1$ so large in this model
gives a physical pion mass above $450\MeV$.

\section{Discussion and outlook}\label{sec:discussion}

In this paper, we have studied the Skyrme model with the addition of
the other possible fourth-order derivative term -- the kinetic term
squared, which however gives rise to more than two time derivatives in
the model; this model has been considered previously in the literature 
(e.g.~\cite{Donoghue:1984yq,Andrianov:1986dn,Lacombe:1985yd,Lacombe:1985mr})
and studied in detail by Pottinger and Rathske
\cite{Pottinger:1985fc}, who performed exact collective coordinate
quantization using Cardano's formula. 

We parametrized the model with a parameter $\eta$ that interpolates
between the Skyrme model ($\eta=1$), the pure kinetic term squared
model ($\eta=1/2$) and a new model with a negative Skyrme term that
completely cancels off the kinetic term squared part ($\eta=0$) and
all the way down to a limit where there is no Skyrme term left
($\eta=-1$), but only a term that vanishes for spherically symmetric
solitons. 
Our result is that the spin contribution only increases once a
four-time derivative term is taken into account, independently of how
many derivatives the term has in total.
This statement holds under the condition of Lorentz-invariant terms
that have a positive definite static energy.
We show this is true by expanding the spin contribution to the energy
in a power series \eqref{eq:Tx-}, where every term is positive.
We further illustrate that this holds for any kind of theory, by
computing the spin contribution to the energy for 4 different
higher-order models.

We further establish topological energy bounds for the models under
consideration in this paper. In particular, we extend the topological
energy bound for the PR model to include two (non-derivative)
potentials, being the pion mass term and the pion mass term squared or
loosely bound potential and we compute new bounds for the higher-order
models. 
These results are given in App.~\ref{app:topobound}.

Unfortunately, it turns out that the ambition of being able to reduce 
the spin contribution to the energy in the class of generalized Skyrme
models with four time derivatives, is not possible.
Hence, the Skyrme model in the class of theories studied in this
paper, is the model with the smallest spin contribution to the energy
and hence the model giving rise to the smallest lower bound on the
binding energy.

The result can have two opposite implications for future work on
achieving physical binding energies. In one direction, one could go
the BPS way and try to reduce the classical binding energies as well
as the spin correction to the energy as much as possible. This is in
the spirit of the assumption that the main contribution to the nucleon
energy is the classical Skyrmion energy, and quantum corrections are
small.
If one chooses to go in another direction of acknowledging that the
classical binding energies should not be small, but only the total
binding energies must sum up to values that are at the percent level
of the mass scale in question, then one could look at this extra
degree of freedom of increasing the spin energy, if needed, as a
tuning parameter. This latter approach to quantization of Skyrmions is
discussed in Ref.~\cite{Gudnason:2023jpq} for the standard Skyrme
model.

An issue with the current model, which we have not solved in this
work, is to perform the quantization for Skyrmions that do not enjoy
spherical symmetry. This becomes complicated because the moment of
inertia tensor will no longer be proportional to the unit matrix; this
implies that the cubic equation that we solve becomes a cubic matrix
equation, that presumably is harder to solve.
One may consider, as a first step, to solve the problem with axial
symmetry, for which two eigenvalues of the tensors are equal.
Nevertheless, if the complete model of quantization of Skyrmions
requires the smallest possible spin contribution, then the Skyrme
model is the best option, to the fourth order in the derivative
expansion.
We will leave for future work, a possible investigation of the spin
contribution for non-spherically symmetric inertia tensors in this
model or other models with more than two time derivatives.
Although more complicated, we believe that the smallest root(s) of the
cubic matrix equation can be found with numerical methods, if it is
not possible to write down analytic expressions.

Another comment in store about theories with four or more time
derivatives, is the problem of the Ostrogradsky instability
\cite{Ostrogradsky:1850fid}.
In the formulation of Woodard \cite{Woodard:2015zca}, the dynamics of
the Hamiltonian is generally described by two or more conjugate momenta, but
only if the Lagrangian is nondegenerate in the double time derivative
of some field.
Luckily, although we have four time derivatives, they each act on
their own field, making the theory highly nonlinear but not inducing
the Ostrogradsky instability.
One may wonder why there is no term like $\tr(U^\dag\square U)^2$ in the most
general Lagrangian of pions, but as shown in the literature such term
can be eliminated by a field redefinition and it will be described at
the four-derivative level just by the two terms included in the
Lagrangian \eqref{eq:Lphysicalunits} plus higher-order terms in the
derivative expansion
\cite{Gasser:1983yg,Gasser:1984gg,Fearing:1994ga,Bijnens:1999sh,Bijnens:2022zqo}.
The Ostrogradsky instability, which exists for nondegenerate double time
derivatives, is due to the fact that the corresponding Hamiltonian will
depend only linearly on one of the two conjugate momenta -- this makes it
possible to drive the theory into larger and larger energies with
either sign.
The theory thus has no lower or upper bound on the
energy.\footnote{
This run-away of the energy is different from the
dynamical instability of the PR model.
That is, in the PR model the instability comes from a negative
contribution to the energy from the four time derivatives, whereas the
Ostrogradsky instability has a linear dependence on one of the
conjugate momenta that can classically cause a run-away at any values
of the kinetic energy. 
}
In the case of the Ostrogradsky instability, arguments have been made
that it is not an issue for EFTs, since the energy needed to excite a
mode that possesses a run-away behavior is larger than the (cutoff)
scale of the EFT and hence anyway beyond the validity of the EFT
\cite{Solomon:2017nlh}. 
It has also been argued that in a certain class of asymptotically free
theories, the effective mass of the unstable modes becomes infinitely
heavy in the UV limit \cite{Asorey:2020omv}.
At some higher order in the derivative expansion, it will no longer be
possible to eliminate all d'Alembertian operators from the EFT
Lagrangian even using field redefinitions and integration-by-parts
relations; at such order the Ostrogradsky instability is inevitable,
although it is possible that it will not cause problems for the EFT
observables at sufficiently small energies.

Finally, one may consider the possibility of going to a higher order
in the derivative expansion.
In the literature, the sextic term has been studied extensively
\cite{Adkins:1983nw,Jackson:1985yz,Adam:2010fg,Adam:2010ds,Gudnason:2016tiz,Gudnason:2018jia},
which however, like the Skyrme term, contains only 2 time derivatives.
A natural generalization of the PR model would be to consider models
with 6 or more time derivatives, which however would give rise to a
higher-order polynomial equation than the cubic equation
\eqref{eq:root_equation}. 
In particular, in the case of theories with 6 time derivatives, the
corresponding order of the polynomial equation is of 5th order.
For a theory with $2n$ time derivatives, the corresponding polynomial
equation for the squared spin operator would then be of order $2n-1$.
In such cases, it is probably necessary to turn to numerical methods
for finding the (smallest/physical) roots.
We leave such problems for future work.

\subsection*{Acknowledgments}

We thank Herbert Weigel for pointing out Ref.~\cite{Pottinger:1985fc}
to us.
S.~B.~G.~thanks Zhang Baiyang for discussions.
S.~B.~G.~thanks the Outstanding Talent Program of Henan University and
the Ministry of Education of Henan Province for partial support.
The work of S.~B.~G.~is supported by the National Natural Science
Foundation of China (Grants No.~11675223 and No.~12071111) and by the
Ministry of Science and Technology of China (Grant No.~G2022026021L).

\appendix

\section{Positivity of static energy}\label{app:positivity_static}

\subsection{The Pottinger-Rathske model}\label{app:positivity_static_PR}

In order to prove the entire suitable range of the couplings while
retaining a positive definite static energy, we rewrite the derivative
part of the static part of the Lagrangian \eqref{eq:V} in terms of the
4-vector field $\bn=(n^0,n^1,n^2,n^3)$: 
\beq
U = n^0\mathbf{1}_2 + \i\tau^an^a, \qquad a=1,2,3,
\label{eq:U_n_vector}
\eeq
yielding
\beq
  \calE &= (\p_i\bn\cdot\p_i\bn) + \frac{2\eta-1}{2}(\p_i\bn\cdot\p_i\bn)^2
  - \frac{2\eta-1}{2}(\p_i\bn\cdot\p_j\bn)^2 + \frac{1-\eta}{2}(\p_i\bn\cdot\p_i\bn)^2.
\eeq
Using the eigenvalues, $\{\lambda_i\}$, of the strain tensor
\cite{Manton:1987xt}
\beq
D_{ij} = -\frac12\tr[R_i R_j]
= (\p_i\bn\cdot\p_j\bn)
= \left[V\begin{pmatrix}\lambda_1^2\\&\lambda_2^2\\&&\lambda_3^2\end{pmatrix}V^{\rm T}\right]_{ij},\qquad
V\in\SO(3),
\eeq
we obtain
\begin{align}
  \calE &= (\lambda_1^2 + \lambda_2^2 + \lambda_3^2)
  +\eta(\lambda_1^2\lambda_2^2 + \lambda_1^2\lambda_3^2 + \lambda_2^2\lambda_3^2)
  +\frac{1-\eta}{2}(\lambda_1^4 + \lambda_2^4 + \lambda_3^4)\non
  &=(\lambda_1^2 + \lambda_2^2 + \lambda_3^2)
  +\frac{1-\eta}{4}\left[(\lambda_1^2-\lambda_2^2)^2+(\lambda_1^2-\lambda_3^2)^2+(\lambda_2^2-\lambda_3^2)^2\right]\non
  &\phantom{=\ }
  +\frac{1+\eta}{2}(\lambda_1^2\lambda_2^2 + \lambda_1^2\lambda_3^2 + \lambda_2^2\lambda_3^2).
  \label{eq:Elambda}
\end{align}
This (derivative part) of the static energy density is thus positive
definite for $\eta\in[-1,1]$.
We expect, however, that the spherically symmetric Skyrmion, which has
$\lambda_1=\lambda_2=\lambda_3$, to be unstable at the point
$\eta=-1$.

\subsection{The higher-order models}\label{app:positivity_static_HO}

In the higher-order models, the static energy is given in
Eq.~\eqref{eq:VHO} with $V'$ of Eq.~\eqref{eq:VHO_8a} for the $8a$
term, Eq.~\eqref{eq:VHO_8b} for the $8b$ term, Eq.~\eqref{eq:VHO_10}
for the $10$ term and Eq.~\eqref{eq:VHO_12} for the $12$ term.
Using the relations in Ref.~\cite{Gudnason:2017opo}, we can write the
derivative part of the static energy density as
\begin{align}
  \calE^{248a} &= (\lambda_1^2 + \lambda_2^2 + \lambda_3^2)
  +\eta(\lambda_1^2\lambda_2^2 + \lambda_1^2\lambda_3^2 + \lambda_2^2\lambda_3^2)
  +\frac{1-\eta}{4}(\lambda_1^2\lambda_2^2 + \lambda_1^2\lambda_3^2 + \lambda_2^2\lambda_3^2)^2,\non
  \calE^{248b} &= (\lambda_1^2 + \lambda_2^2 + \lambda_3^2)
  +\eta(\lambda_1^2\lambda_2^2 + \lambda_1^2\lambda_3^2 + \lambda_2^2\lambda_3^2)
  +\frac{1-\eta}{4}(\lambda_1^2 + \lambda_2^2 + \lambda_3^2)(\lambda_1^2\lambda_2^2\lambda_3^2),\non
  \calE^{24(10)} &= (\lambda_1^2 + \lambda_2^2 + \lambda_3^2)
  +\eta(\lambda_1^2\lambda_2^2 + \lambda_1^2\lambda_3^2 + \lambda_2^2\lambda_3^2)
  +\frac{1-\eta}{4}(\lambda_1^2\lambda_2^2 + \lambda_1^2\lambda_3^2 + \lambda_2^2\lambda_3^2)(\lambda_1^2\lambda_2^2\lambda_3^2),\non
  \calE^{24(12)} &= (\lambda_1^2 + \lambda_2^2 + \lambda_3^2)
  +\eta(\lambda_1^2\lambda_2^2 + \lambda_1^2\lambda_3^2 + \lambda_2^2\lambda_3^2)
  +\frac{1-\eta}{4}(\lambda_1^4\lambda_2^4\lambda_3^4).
\end{align}
Positivity of all these static energy functionals can only be guaranteed if
$\eta\in[0,1]$.

For the $8a$ case, we can see this by considering a very small
Skyrme-energy contribution, for which the last term is negligible.
This requires $\eta\geq0$.
For regions with a large Skyrme-energy contribution, the last term is
dominant and we need $1-\eta\geq 0$.

For the $8b$, $10$ and $12$ cases, one can consider regions in
space where the baryon density vanishes, but two of the
strain tensor eigenvalues do not (i.e.~$\lambda_1\neq0$,
$\lambda_2\neq0$ and $\lambda_3=0$). In this case, the last term
vanishes and $\eta\geq0$ is a necessity, hence in general it is.

\section{Topological energy bound}\label{app:topobound}

\subsection{The Pottinger-Rathske model}\label{app:topobound_PR}

Let us consider the topological energy bound for the derivative terms
of the static energy \eqref{eq:V}.
  Using the second line of Eq.~\eqref{eq:Elambda}, we can write the
topological bound on the static energy \cite{Pottinger:1985fc}
\begin{align}
  E &= \int\d^3x\bigg[
    (\lambda_1^2 + \lambda_2^2 + \lambda_3^2)
  +\frac{1-\eta}{4}\left[(\lambda_1^2-\lambda_2^2)^2+(\lambda_1^2-\lambda_3^2)^2+(\lambda_2^2-\lambda_3^2)^2\right]\non
  &\phantom{=\int\d^3x\bigg[\ }
  +\frac{1+\eta}{2}(\lambda_1^2\lambda_2^2 + \lambda_1^2\lambda_3^2 + \lambda_2^2\lambda_3^2)\bigg]\non
  & =\int\d^3x\bigg[
    \bigg(\lambda_1 \mp \sqrt{\frac{1+\eta}{2}}\lambda_2\lambda_3\bigg)^2
    +\bigg(\lambda_2 \mp \sqrt{\frac{1+\eta}{2}}\lambda_3\lambda_1\bigg)^2
    +\bigg(\lambda_3 \mp \sqrt{\frac{1+\eta}{2}}\lambda_1\lambda_2\bigg)^2\non
    &\phantom{=\int\d^3x\bigg[\ }
    +\frac{1-\eta}{4}\left[(\lambda_1^2-\lambda_2^2)^2+(\lambda_1^2-\lambda_3^2)^2+(\lambda_2^2-\lambda_3^2)^2\right]
    \pm6\sqrt{\frac{1+\eta}{2}}\lambda_1\lambda_2\lambda_3\bigg]\non
  &\geq6\sqrt{\frac{1+\eta}{2}}\int\d^3x\;|\lambda_1\lambda_2\lambda_3|\non
  &\geq12\pi^2\sqrt{\frac{1+\eta}{2}}|B|,
\end{align}
where $B$ is the baryon number of Eq.~\eqref{eq:B} and we have used
the fact that $\int\d^3x\;\lambda_1\lambda_2\lambda_3=2\pi^2B$.
We can see that the Skyrme bound on the energy ($12\pi^2|B|$)
\cite{Skyrme:1962vh} is recovered for $\eta=1$ and that the bound goes
to zero for $\eta\to-1$, signaling a possible instability for the
spherically symmetric case.

Taking into account the non-derivative part of the potential energy can
be done following Harland \cite{Harland:2013rxa}, see also
Refs.~\cite{Adam:2013tga,Gudnason:2022jkn}.
We define the following functions
\begin{align}
  E_2 &= \int\d^3x\;(\lambda_1^2 + \lambda_2^2 + \lambda_3^2),\label{eq:E2}\\
  E_4 &= \int\d^3x\;(\lambda_1^2\lambda_2^2 + \lambda_1^2\lambda_3^2 + \lambda_2^2\lambda_3^2),\label{eq:E4}\\
  E_{01} &= \frac12\int\d^3x\;\tr(\mathbf{1}_2-U),\label{eq:E01}\\
  E_{02} &= \frac14\int\d^3x\;\left[\tr(\mathbf{1}_2-U)\right]^2.\label{eq:E02}
\end{align}
We write the maximization problem as
\begin{align}
  E &= \left(E_2 + \alpha\frac{1+\eta}{2}E_4\right)
  + \left(2m_1^2E_{01} + \beta(1-\alpha)\frac{1+\eta}{2}E_4\right)\non
  &\phantom{=\ }
  + \left(m_2^2E_{02} +
  (1-\beta)(1-\alpha)\frac{1+\eta}{2}E_4\right)\non
  &\geq12\pi^2\bigg[\sqrt{\frac{\alpha(1+\eta)}{2}}
    +\frac{128\sqrt{m_1}\beta^{\frac34}(1-\alpha)^{\frac34}(1+\eta)^{\frac34}\Gamma^2\left(\tfrac34\right)}{45\times2^{\frac34}\pi^{\frac32}}\non
    &\phantom{\geq12\pi^2\bigg[\ }
    +\frac{64\sqrt{m_2}(1-\beta)^{\frac34}(1-\alpha)^{\frac34}(1+\eta)^{\frac34}}{45\pi}
    \bigg]|B|,
  \label{eq:E_PR_derivative_bound}
\end{align}
with $\alpha\in[0,1]$ and $\beta\in[0,1]$, which is a maximization
problem in two real variables on the unit interval,
where we have used the bound \cite{Harland:2013rxa}:
\beq
2m_1^2\int\d^3x\;v(\tr U) + \mu E_4
\geq 16\pi\times 2^{\frac14}\sqrt{m_1}\mu^{\frac34}|B|\int_0^\pi\sin^2(f)\left[v(2\cos f)\right]^{\frac14}\;\d f.
\eeq
The result of the maximization of Eq.~\eqref{eq:E_PR_derivative_bound}
does not only depend on the masses $m_1$ and $m_2$, but also on the
chosen value of $\eta$.
First we extremize with respect to $\beta$, obtaining
\begin{align}
  E &\geq12\pi^2\left[\sqrt{\frac{\alpha(1+\eta)}{2}}
    +\frac{64(1-\alpha)^{\frac34}(1+\eta)^{\frac34}}{45\pi}\left(2^{\frac14}\sqrt{\frac{m_1}{\pi}}\Gamma^2\left(\tfrac34\right)F(\zeta)+\sqrt{m_2}F(\zeta^{-1})\right)
    \right]|B|,
\end{align}
with
\beq
F(\zeta) = \left(\frac{1}{1+\zeta}\right)^{\frac34}, \qquad
\zeta = \frac{\pi^2m_2^2}{2m_1^2\Gamma^8\left(\tfrac34\right)}.
\eeq
On the other hand, maximization with respect to $\alpha$ yields
\begin{equation}
\alpha = \frac{a^2}{2}\left(\sqrt{1+\frac{4}{a^2}}-1\right),\qquad
a = \frac{225\pi^2(1+\eta)}{2048(1+\eta)^{\frac32}\left(2^{\frac14}\sqrt{\frac{m_1}{\pi}}\Gamma^2\left(\tfrac34\right)F(\zeta)+\sqrt{m_2}F(\zeta^{-1})\right)^2},
\label{eq:alpha_solution}
\end{equation}
which is valid for $\eta\in(-1,1]$.

\subsection{The higher-order models}\label{app:topobound_HO}

We start by considering the derivative part of the higher-order
models, which have the terms \eqref{eq:VHO_8a}-\eqref{eq:VHO_12}. 
We start by determining new bounds on higher-order terms, that have
not been derived earlier, to the best of our knowledge.
Starting with the term \eqref{eq:VHO_8a}, we write
\begin{align}
  E &= \int\d^3x\left[\mu(\lambda_1^2\lambda_2^2 +
    \lambda_1^2\lambda_3^2 + \lambda_2^2\lambda_3^2)^2 + 2m_1^2 v(\tr
    U)\right]\non
  &= \frac38\left[\frac83\mu\int\d^3x\;(\lambda_1^2\lambda_2^2 +
      \lambda_1^2\lambda_3^2 + \lambda_2^2\lambda_3^2)^2\right]
  + \frac58\left[\frac85\int\d^3x\;2m_1^2 v(\tr U)\right]\non
  &\geq\left[\frac83\mu\int\d^3x\;(\lambda_1^2\lambda_2^2 +
      \lambda_1^2\lambda_3^2 + \lambda_2^2\lambda_3^2)^2\right]^{\frac38}
  \left[\frac85\int\d^3x\;2m_1^2 v(\tr U)\right]^{\frac58},
  \label{eq:topobound_HO_first_step}
\end{align}
where we have used the inequality of the arithmetic and geometric means
\beq
\sum_{a=1}^n w_a x_a \geq \prod_{a=1}^n x_a^{w_a},
\eeq
which holds for non-negative $x_a$ and $w_1+w_2+\cdots w_n=1$, all
positive as well.
Using the same inequality again, we have \cite{Harland:2013rxa}
\beq
\frac13(\lambda_1^2\lambda_2^2 + \lambda_1^2\lambda_3^2 + \lambda_2^2\lambda_3^2)
\geq |\lambda_1\lambda_2\lambda_2|^{\frac43},
\label{eq:E4_B_inequality}
\eeq
and consequently
\beq
\frac19(\lambda_1^2\lambda_2^2 + \lambda_1^2\lambda_3^2 + \lambda_2^2\lambda_3^2)^2
\geq |\lambda_1\lambda_2\lambda_2|^{\frac83}.
\eeq
We can now write
\begin{align}
  E &\geq8\left[3\mu\int\d^3x\;|\lambda_1\lambda_2\lambda_3|^{\frac83}\right]^{\frac38}
  \left[\frac15\int\d^3x\;2m_1^2 v(\tr U)\right]^{\frac58},
  \label{eq:E08a_bound}
\end{align}
where everything has been chosen carefully such that the power
$\frac38$ is the inverse of $\frac83$.

At this point, it will prove convenient to write the expression with a
positive integer $n$ as
\begin{align}
  E &\geq n\left[\frac{\mu'}{3}\int\d^3x\;|\lambda_1\lambda_2\lambda_3|^{\frac{n}{3}}\right]^{\frac3n}
  \left[\frac{1}{n-3}\int\d^3x\;2m_1^2 v(\tr U)\right]^{\frac{n-3}{n}},\qquad n>3.
\end{align}
Now we utilize the H\"older's inequality ($\tfrac1p+\tfrac1q=1$, $p>0$, $q>0$):
\beq
\left(\int\d^3x\;|f_1|^p\right)^{\frac1p}
\left(\int\d^3x\;|f_2|^q\right)^{\frac1q}
\geq\int\d^3x\;|f_1f_2|,
\eeq
to obtain
\begin{align}
  E &\geq n\left(\frac{\mu'}{3}\right)^{\frac3n}\left(\frac{2m_1^2}{n-3}\right)^{\frac{n-3}{n}}
  \int\d^3x\;|v(\tr U)|^{\frac{n-3}{n}}|\lambda_1\lambda_2\lambda_3|,
\end{align}
which can be written in terms of the topological degree $B$ as
\begin{align}
  E &\geq
  4n\pi\left(\frac{\mu'}{3}\right)^{\frac3n}\left(\frac{2m_1^2}{n-3}\right)^{\frac{n-3}{n}}|B|
  \int_0^\pi\sin^2(f)|v(2\cos f)|^{\frac{n-3}{n}}\d f.
  \label{eq:topobound_n}
\end{align}
In particular, for the case of Eq.~\eqref{eq:E08a_bound} with $n=8$
and $\mu'=9\mu$, we have for the pion mass term
\begin{align}
  \int\d^3x\left[\mu(\lambda_1^2\lambda_2^2 +\lambda_1^2\lambda_3^2 + \lambda_2^2\lambda_3^2)^2 + m_1^2\tr(\mathbf{1}_2-U)\right]
  \geq12\pi^2
  \frac{256\sqrt{\pi}2^{\frac14}(3\mu)^{\frac38}\big(\frac{m_1^2}{5}\big)^{\frac58}}
  {455\sin\big(\tfrac{\pi}{8}\big)\Gamma\big(\tfrac58\big)\Gamma\big(\tfrac78\big)}|B|.
\end{align}

We will now turn to the term \eqref{eq:VHO_8b}:
\begin{align}
  E &= \int\d^3x\left[\mu(\lambda_1^2 + \lambda_2^2 + \lambda_3^2)
    (\lambda_1^2\lambda_2^2\lambda_3^2) + 2m_1^2 v(\tr U)\right]\non
  &\geq8\left[\frac\mu3\int\d^3x\;(\lambda_1^2 + \lambda_2^2 + \lambda_3^2)
    (\lambda_1^2\lambda_2^2\lambda_3^2)\right]^{\frac38}
  \left[\frac15\int\d^3x\;2m_1^2 v(\tr U)\right]^{\frac58}\non
  &\geq8\left[\mu\int\d^3x\;
    |\lambda_1\lambda_2\lambda_3|^{\frac83}\right]^{\frac38}
  \left[\frac15\int\d^3x\;2m_1^2 v(\tr U)\right]^{\frac58}\non
  &\geq 32\pi\mu^{\frac38}\left(\frac{2m_1^2}{5}\right)^{\frac{5}{8}}|B|
  \int_0^\pi\sin^2(f)|v(2\cos f)|^{\frac{5}{8}}\d f,
\end{align}
where we have used the step of Eq.~\eqref{eq:topobound_HO_first_step},
the inequality
\beq
\frac13(\lambda_1^2+\lambda_2^2+\lambda_3^2)\geq|\lambda_1\lambda_2\lambda_3|^{\frac23},
\eeq
as well as Eq.~\eqref{eq:topobound_n}.
Using the pion mass term as the non-derivative potential, we get
\begin{equation}
\int\d^3x\left[\mu(\lambda_1^2 + \lambda_2^2 + \lambda_3^2)
    (\lambda_1^2\lambda_2^2\lambda_3^2) + m_1^2\tr(\mathbf{1}_2-U)\right]
\geq12\pi^2
\frac{256\sqrt{\pi}2^{\frac14}\mu^{\frac38}\big(\tfrac{m_1^2}{5}\big)^{\frac58}}{455\sin\big(\tfrac\pi8\big)\Gamma\big(\tfrac58\big)\Gamma\big(\tfrac78\big)}|B|.
\end{equation}

The next higher-order term is \eqref{eq:VHO_10}, which is a 10th order
derivative term:
\begin{align}
  E &= \int\d^3x\left[\mu(\lambda_1^2\lambda_2^2 + \lambda_1^2\lambda_3^2 + \lambda_2^2\lambda_3^2)
    (\lambda_1^2\lambda_2^2\lambda_3^2) + 2m_1^2 v(\tr U)\right]\non
  &\geq10\left[\frac\mu3\int\d^3x\;(\lambda_1^2\lambda_2^2 + \lambda_1^2\lambda_3^2 + \lambda_2^2\lambda_3^2)
    (\lambda_1^2\lambda_2^2\lambda_3^2)\right]^{\frac{3}{10}}
  \left[\frac17\int\d^3x\;2m_1^2 v(\tr U)\right]^{\frac{7}{10}}\non
  &\geq10\left[\mu\int\d^3x\;
    |\lambda_1\lambda_2\lambda_3|^{\frac{10}{3}}\right]^{\frac{3}{10}}
  \left[\frac17\int\d^3x\;2m_1^2 v(\tr U)\right]^{\frac{7}{10}}\non
  &\geq 40\pi\mu^{\frac{3}{10}}\left(\frac{2m_1^2}{7}\right)^{\frac{7}{10}}|B|
  \int_0^\pi\sin^2(f)|v(2\cos f)|^{\frac{7}{10}}\d f,
\end{align}
where we have used the step of Eq.~\eqref{eq:topobound_HO_first_step},
the inequality \eqref{eq:E4_B_inequality} as well as Eq.~\eqref{eq:topobound_n}.
Using the pion mass term as the non-derivative potential, we get
\begin{align}
\int\d^3x\left[\mu(\lambda_1^2\lambda_2^2 + \lambda_1^2\lambda_3^2 + \lambda_2^2\lambda_3^2)
    (\lambda_1^2\lambda_2^2\lambda_3^2) + m_1^2\tr(\mathbf{1}_2-U)\right]\qquad\qquad\qquad\non
\geq12\pi^2
\frac{640\sqrt{5(5+\sqrt{5})\pi}\,(8\mu)^{\frac{3}{10}}\big(\tfrac{m_1^2}{7}\big)^{\frac{7}{10}}}{3213\,\Gamma\big(\tfrac{7}{10}\big)\Gamma\big(\tfrac45\big)}|B|.
\end{align}

The last higher-order term is \eqref{eq:VHO_12}, which is a 12th order
derivative term:
\begin{align}
  E &= \int\d^3x\left[\mu(\lambda_1^4\lambda_2^4\lambda_3^4) + 2m_1^2 v(\tr U)\right]\non
  &\geq4\left[\mu\int\d^3x\;
    |\lambda_1\lambda_2\lambda_3|^4\right]^{\frac14}
  \left[\frac13\int\d^3x\;2m_1^2 v(\tr U)\right]^{\frac34}\non
  &\geq 16\pi\mu^{\frac14}\left(\frac{2m_1^2}{3}\right)^{\frac34}|B|
  \int_0^\pi\sin^2(f)|v(2\cos f)|^{\frac34}\d f,
\end{align}
where we have used the step of Eq.~\eqref{eq:topobound_HO_first_step}
and Eq.~\eqref{eq:topobound_n}.
Using the pion mass term as the non-derivative potential, we get
\begin{equation}
\int\d^3x\left[\mu(\lambda_1^4\lambda_2^4\lambda_3^4) + m_1^2\tr(\mathbf{1}_2-U)\right]
\geq12\pi^2
\frac{1280\mu^{\frac14}\big(\tfrac{m_1^2}{3}\big)^{\frac34}K\big(\tfrac12\big)}{693\pi}|B|,
\end{equation}
where $K(m)$ is the complete elliptic integral of the first kind and
$K(0.5)\approx1.85407$. 

We will now write the complete topological energy bounds for the
higher-order models, using the new results for the subbounds that
involve the higher-order derivative term and a potential term.
For simplicity, we will turn on only the pion mass term here, and
switch off the other non-derivative potential, i.e.~setting $m_2:=0$.
Writing the higher-order model \eqref{eq:VHO} as
\beq
E_{\rm HO}^{(X)} = (E_2 + \alpha\eta E_4)
+ (2\beta m_1^2E_{01} + (1-\alpha)\eta E_4)
+ \left(2(1-\beta)m_1^2E_{01} + \frac{1-\eta}{4}E_{X}\right),
\eeq
where $E_{01}$, $E_2$ and $E_4$ are given by Eqs.~\eqref{eq:E01},
\eqref{eq:E2} and \eqref{eq:E4}, respectively, and $E_X$ is the static
energy of a higher-order derivative term.
We can now write the full topological energy bounds for all the
higher-order models as
\begin{align}
  E_{\rm HO}^{(8a)}
  &\geq12\pi^2\Bigg[
    \sqrt{\alpha\eta}
    +\frac{128\beta^{\frac14}\sqrt{m_1}\left((1-\alpha)\eta\right)^{\frac34}\Gamma^2\big(\tfrac34\big)}{45\pi^{\frac32}}
    +\frac{128\sqrt{2\pi}
    \left(3(1-\eta)\right)^{\frac38}\left(\frac{(1-\beta)m_1^2}{5}\right)^{\frac58}}
  {455\sin\big(\tfrac{\pi}{8}\big)\Gamma\big(\tfrac58\big)\Gamma\big(\tfrac78\big)}
  \Bigg]|B|,\non
  E_{\rm HO}^{(8b)}
  &\geq12\pi^2\Bigg[
    \sqrt{\alpha\eta}
    +\frac{128\beta^{\frac14}\sqrt{m_1}\left((1-\alpha)\eta\right)^{\frac34}\Gamma^2\big(\tfrac34\big)}{45\pi^{\frac32}}
    +\frac{128\sqrt{2\pi}
    \left((1-\eta)\right)^{\frac38}\left(\frac{(1-\beta)m_1^2}{5}\right)^{\frac58}}
  {455\sin\big(\tfrac{\pi}{8}\big)\Gamma\big(\tfrac58\big)\Gamma\big(\tfrac78\big)}
  \Bigg]|B|,\non
  E_{\rm HO}^{(10)}
  &\geq12\pi^2\Bigg[
    \sqrt{\alpha\eta}
    +\frac{128\beta^{\frac14}\sqrt{m_1}\left((1-\alpha)\eta\right)^{\frac34}\Gamma^2\big(\tfrac34\big)}{45\pi^{\frac32}}\non
    &\phantom{\geq12\pi^2\Bigg[\ }
    +\frac{640\sqrt{5(5+\sqrt{5})\pi}
    \left(2(1-\eta)\right)^{\frac{3}{10}}\left(\frac{(1-\beta)m_1^2}{7}\right)^{\frac{7}{10}}}
  {3213\Gamma\big(\tfrac{7}{10}\big)\Gamma\big(\tfrac45\big)}
  \Bigg]|B|,\non
  E_{\rm HO}^{(12)}
  &\geq12\pi^2\Bigg[
    \sqrt{\alpha\eta}
    +\frac{128\beta^{\frac14}\sqrt{m_1}\left((1-\alpha)\eta\right)^{\frac34}\Gamma^2\big(\tfrac34\big)}{45\pi^{\frac32}}
    +\frac{640(1-\eta)^{\frac14}\left(\frac{(1-\beta)m_1^2}{3}\right)^{\frac34}}{693\pi}
    \Bigg]|B|,
  \label{eq:E024X_HO_topobounds}
\end{align}
and we have defined
\begin{align}
E_{8a} &= \int\d^3x\;(\lambda_1^2\lambda_2^2 +\lambda_1^2\lambda_3^2 + \lambda_2^2\lambda_3^2)^2,\non
E_{8b} &= \int\d^3x\;(\lambda_1^2 + \lambda_2^2 + \lambda_3^2)(\lambda_1^2\lambda_2^2\lambda_3^2),\non
E_{10} &= \int\d^3x\;(\lambda_1^2\lambda_2^2 + \lambda_1^2\lambda_3^2 + \lambda_2^2\lambda_3^2)(\lambda_1^2\lambda_2^2\lambda_3^2),\non
E_{12} &= \int\d^3x\;(\lambda_1^4\lambda_2^4\lambda_3^4).
\end{align}
The maximization of the bounds \eqref{eq:E024X_HO_topobounds} with
respect to $\alpha$ can be carried out easily as in the
Pottinger-Rathske model (see App.~\eqref{app:topobound_PR}) if it is
done first, but since the $(1-\alpha)$-dependence is only included in
the Skyrme term and not in the higher-order term, the extremization
with respect to $\alpha$ is changed and possibly becomes more
difficult after maximization has been done with respect to $\beta$.

Starting with the $8a$ and $8b$ cases, extremization with respect to
$\beta$ can be carried out as:
\begin{align}
\beta &=
\frac{2}{25}\left(\frac25\right)^{\frac23}\xi^{\frac83}
\left[\sqrt{1 + 25\left(\frac52\right)^{\frac23}\xi^{-\frac83}} - 1\right],\\
\xi &= \frac{91\sqrt{m_1}((1-\alpha)\eta)^{\frac34}\Gamma\big(\tfrac58\big)\Gamma^2\big(\tfrac34\big)\Gamma\big(\tfrac78\big)\sin\big(\tfrac\pi8\big)}{9\sqrt{2}\pi^2\big(\frac{(1-\beta)m_1^2}{5}\big)^{\frac58}(3(1-\eta))^{\frac38}}.
\end{align}
We are, however, unable to perform the maximization with respect to
$\alpha$ analytically at this point.
For the $10$ case, the extremization with respect to $\beta$ requires
the root to 5th order polynomial equation, that we do not know in
closed form.
For the $12$ case, maximization with respect to $\beta$ is given by
a real root to a 3rd order polynomial equation, which can be found by
Cardano's formula.
However, we are also in this case unable to perform the analytic
maximization with respect to $\alpha$ after having maximized with
respect to $\beta$.

We are thus not able to find analytic expressions for both $\alpha$
and $\beta$ in the cases of the higher-order models that
maximize their respective bounds, but the maximization in the two
variables $\alpha$ and $\beta$ on the unit interval can for fixed
values of $m_1$ and $\eta$ easily be done numerically.

Taking into account both non-derivative potentials ($m_1>0$ and
$m_2>0$) can straightforwardly be done as in the Pottinger-Rathske
model by introducing a third parameter (see
App.~\ref{app:topobound_PR}), which we will leave as an exercise.
Again only a numerical solution to the three parameters will be
possible.

\bibliographystyle{JHEP}
\bibliography{references.bib}

\end{document}